\documentclass[amssymb,amsmath,11pt]{article}
\usepackage{natbib,graphicx,subfigure,subeqnarray,fancyhdr,amssymb,amsmath,epstopdf,geometry,setspace,mathrsfs,color,multirow}
\begin{document}

\newcommand{\Sc}{\mathscr{S}}
\newcommand{\Fc}{\mathcal{F}}\newcommand{\Rc}{\mathcal{R}}\newcommand{\dd}{\mathrm{d}}
\newcommand{\ee}{\mathrm{e}}\newcommand{\ci}{\mathrm{i}}\newcommand{\ib}{\mathbf{i}}
\newcommand{\jb}{\mathbf{j}}\newcommand{\kb}{\mathbf{k}}\newcommand{\ab}{\mathbf{a}}
\newcommand{\Fb}{\mathbf{F}}\newcommand{\fb}{\mathbf{f}}\newcommand{\Gb}{\mathbf{G}}
\newcommand{\Mb}{\mathbf{M}Ä}\newcommand{\nb}{\mathbf{n}}\newcommand{\Sb}{\mathbf{S}}
\newcommand{\Sbs}{\mathbf{S^*}}\newcommand{\Rb}{\mathbf{R}}\newcommand{\Sigb}{\boldsymbol{\Sigma}}
\newcommand{\Sigbs}{\boldsymbol{\Sigma^*}}\newcommand{\omegab}{\boldsymbol{\omega}}\newcommand{\epsb}{\boldsymbol{\epsilon}}
\newcommand{\ub}{\mathbf{u}}
\newcommand{\eb}{\mathbf{e}}\newcommand{\vv}[1]{\underline{#1}}\newcommand{\ev}{\vv{e}}
\newcommand{\rv}{\vv{r}}\newcommand{\TT}[1]{\underline{\underline{#1}}}\newcommand{\omb}{\mathbf{\omega}}
\newcommand{\Ub}{\mathbf{U}}\newcommand{\xb}{\mathbf{x}}\newcommand{\rb}{\mathbf{r}}
\newcommand{\ssb}{\mathbf{s}}\newcommand{\Xb}{\mathbf{X}}\newcommand{\Rey}{\mbox{\textit{Re}}}
\newcommand{\ddp}{[p]^\pm}\newcommand{\taub}{\mbox{\boldmath$\tau$}}\newcommand{\Fr}{\mbox{\textit{Fr}}}
\let\grad\nabla\newcommand{\z}{\zeta}\newcommand{\kk}{\kappa}\newcommand{\tkk}{\tilde{\kappa}}
\newcommand{\e}{\varepsilon}\newcommand{\zb}{\bar{\zeta}}\let\grad\nabla\let\bcdot\cdot
\newcommand{\half}{{\textstyle\frac{1}{2}}}
\newcommand{\textfrac}[2]{{\textstyle\frac{#1}{#2}}}
\newcommand{\LF}[1]{{#1}^{\mathrm{LF}}}\newcommand{\Lap}[1]{{#1}^{\mathrm{L}}}
\newcommand{\ds}{*\!*}\newcommand{\cond}[2]{\frac{\mathrm{D} #1}{\mathrm{D} #2}}
\newcommand{\pard}[2]{\frac{\partial #1}{\partial #2}}\newcommand{\totd}[2]{\frac{\mathrm{d}#1}{\mathrm{d}#2}}
\newcommand{\Real}{\mbox{Re}}\newcommand{\Imag}{\mbox{Im}}
\newcommand{\Fpint}{=\!\!\!\!\!\!\!\int}
\makeatletter
\def\sgn{\mathop{\operator@font sgn}}
\makeatother

\title{{\bf The long-time dynamics of two  hydrodynamically-coupled swimming cells}}
\author{S\'ebastien Michelin\footnote{smichelin@ucsd.edu}~  and Eric Lauga\footnote{elauga@ucsd.edu}\\
\small{\emph{Department of Mechanical and Aerospace Engineering, University of California San Diego,}}\\ \small{\emph{ 9500 Gilman Drive, La Jolla CA 92093-0411.}}}
%\email{smichelin@ucsd.edu}
%\author{Eric Lauga}
%\email{elauga@ucsd.edu}
%\affiliation{Department of Mechanical and Aerospace Engineering, University of California San Diego, 9500 Gilman Drive, La Jolla CA 92093-0411.}
\date{\today}
\maketitle
\begin{abstract}

Swimming micro-organisms such as bacteria or spermatozoa are typically found in dense suspensions, and exhibit collective modes of locomotion qualitatively different from that displayed by isolated cells. In the dilute limit where fluid-mediated interactions can be treated rigorously, the long-time hydrodynamics of a collection of cells result from interactions with many other cells, and as such typically eludes an analytical approach. Here we consider the only case where such problem can be treated rigorously analytically, namely when the cells have spatially confined trajectories, such as the spermatozoa of some marine invertebrates. We consider two spherical cells swimming, when isolated, with arbitrary circular trajectories, and derive the long-time  kinematics of their relative locomotion. We show that in the dilute limit where the cells are much further away than their size, and the size of their circular motion, a separation of time scale occurs between a fast (intrinsic) swimming time, and a slow time where hydrodynamic interactions lead to change in the relative position and orientation of the swimmers. We perform a multiple-scale analysis and derive the effective dynamical system --- of dimension two --- describing the long-time behavior of the pair of cells. We show that the system displays one type of equilibrium, and two types of rotational equilibrium, all of which are found to be unstable. A detailed mathematical analysis of the dynamical systems further allows us to show that only two cell-cell behaviors are possible in the limit of $t\to\infty$, either the cells are attracted to each other (possibly monotonically), or they are repelled (possibly monotonically as well), which we confirm with numerical computations. Our analysis shows therefore that, even in the dilute limit, hydrodynamic interactions lead to new modes of  cell-cell locomotion.\\


\end{abstract}

\doublespacing
\section{Introduction}

Micro-organisms such as bacteria and simple eukaryotes are found in nature in a variety of environments, from large water masses (ocean, lakes, rivers) to the fluid components of plants and animals. In all, they represent half of the world's biomass, and have therefore major biological consequences on the health and survival of most other organisms.

When a micro-organism has the ability to swim in a viscous fluid, then its motion is the complicated result of the local transport by the moving  fluid it resides in, and of its intrinsic swimming. Given the small size, $\ell$, of these micro-organisms (typically $\ell\approx 1$--$10$ $\mu$m) and the small swimming velocities,  $V$ (typically $V\approx 10-100$ $\mu$m/s), the Reynolds number, $\Rey=V\ell/\nu$, is much smaller than $1$ (here $\nu$ is the kinematic viscosity of the fluid). For such swimmers, the interactions with the surrounding fluid are therefore dominated by viscous stresses, and inertial effects are negligible \citep{lighthill75}. As a results, the velocity and pressure fields around the swimmer  satisfy Stokes' equations  \citep{happel,kimbook}.

Most classical work on the dynamics of swimming cells considered the mechanics and physics of individual organisms \citep{lighthill76,brennen77,blum79,childress81,laugapowers,braybook}. However, cells are typically found in large dense suspensions, and display collective modes of locomotion which are qualitatively different from that of individual cells. For example spermatozoa populations can be as large as millions, and in some species display aggregation and cooperative locomotion. Such is the case for wood mouse spermatozoa \citep{moore02}, as well as opossum \citep{moore95} and fishfly \citep{hayashi98}. Concentrated bacterial suspensions display large-scale coherent and intermittent collective swimming, with length and velocity scales much larger than that of a single cell \citep{mendelson99,dombrowski04,sokolov07,cisneros07}, and resulting in an enhanced diffusion of suspended particles \citep{wu00,kim04_diffusion}.

Significant work has been devoted to the theoretical modeling of collective effects in cell locomotion. Building on early work showing that dipole-dipole hydrodynamic interactions between swimming cells lead to aggregation \citep{guell88}, two distinct approaches have been  considered. On one hand, continuum studies have been proposed in the dilute limit. Classical work on bioconvection neglected the presence of swimming cells altogether \citep{childress75,pedley92,hill05}. When the swimmer size is small compared to the typical inter-swimmer distance, the first effect of a self-propelled micro-organism is to modify the local stresses in the flow by creating a local  dipolar (or stresslet) forcing on the surrounding fluid  \citep{batchelor1970}. Within this framework, studies have discovered long-wavelength hydrodynamic instabilities occurring in suspensions of self-propelled bodies \citep{simha02,saintillan08}. The resulting nonlinear state, sometimes referred to as ``bacterial turbulence''  has also been reproduced using continuum simulations \citep{aranon07,wolgemuth08}. On the other hand, a number of studies have focused on the discrete nature of the ``N-swimming body'' problem, and solved numerically for the dynamics of each self-propeled body. Models of increasing complexity have represented the swimmer as a point-dipole  \citep{hernandez-ortiz05,underhill08}, a line distribution of surface stress \citep{saintillan07}, or a surface distribution of tangential velocity  \citep{ishikawa_pedley_rheology07,ishikawa_pedley_diffusion07,ishikawa08}, and have reproduced some of  the instabilities, diffusive behavior, and nonlinear dynamics  observed experimentally  \citep[see also][]{mehandia08}.  The subtle role of hydrodynamic interactions in allowing for new modes of locomotion was also recently pointed out  \citep{alexander08,laugabartolo08}. In parallel, work in the physics community has discovered phase-transitions to collective motion in kinematics models of large populations of self-propelled bodies without the need for hydrodynamic interactions \citep{vicsek95,czirok97,gregoire04}.

From a theoretical standpoint, collective locomotion is a difficult problem.  To be treated satisfactorily, the motion of $N\gg 1$ identical swimmers should be integrated in time. In the dense limit, no simple model is available to correctly describe the interplay between hydrodynamic and steric (excluded-volume) interactions. One simplification is to consider the dilute limit, in which hydrodynamic interactions can be described by dipole-dipole interactions, and steric interactions can be neglected. However in this limit, hydrodynamic interactions are weak, and an order-one change in the trajectory of a straight-swimming body can only result from a large number of successive interactions with different swimmers. In other words, even in the dilute limit, one needs in general to study $N\gg1$ cells to quantitatively capture their coupled dynamics. 

In this paper, we consider the only situation in which the case of $N=2$ swimmers can give rise to order-one changes in the long-time limit of their positions and orientations even in the dilute limit, namely when the individual swimmers have {\it spatially confined} intrinsic trajectories.  In that case, even small hydrodynamic interaction can accumulate over times long compared to an intrinsic swimming time, and lead to nontrivial nonlinear dynamics of the coupled system.
By studying in the long-time limit one of these prototypical situations, we hope to obtain important physical and mathematical insight on the general  behavior of larger populations.

We focus our study on the particular situation where the intrinsic motion of the micro-organisms is circular. This is the case, for example, for sea urchin spermatozoa \citep{riedel05}, or other marine invertebrates \citep{goldstein77}. We consider two identical but arbitrary  model cells, and assume they are widely separated. This  assumption allows us to propose a simple general representation of cell-cell hydrodynamic interactions in  \S \ref{sec:geneqmotion}. We then show that a separation of time scales occurs, with a short time representing  the intrinsic swimming time for each cell, and the long time being the time one has to wait for repeated hydrodynamic interactions to lead to order-one changes in the swimmers  trajectories. This separation of time scales allows us to perform  a multiple-scale analysis of the coupled dynamics in \S \ref{multscale}. The equilibrium configurations of the two cells, as well as their stability, are studied in \S \ref{sec:dynsys}. The time-averaged equations are reduced to a two-dimensional dynamical system whose behavior is analyzed in detail. In particular, we show that only two long-time behaviors can arise, as determined solely by the initial relative orientations of the swimmers: Either hydrodynamic interactions have a net repulsive effect and the swimmers eventually swim infinitely far away from each other, or they have a net attractive effect, and lead to  collisions (or aggregation) of the two swimmers. Any relative equilibrium or limit cycle is found to be unstable, and we therefore do not observe any organization of the swimmers' motion through hydrodynamic interactions. Our modeling assumptions, some possible extensions, and the relevance to biological locomotion 
 are discussed in \S\ref{discussion}.

%% SECTION 2 GENERAL EQUATIONS
\section{Equations of motion of two Stokesian swimmers}
\label{sec:geneqmotion}
\subsection{Intrinsic motion}
\begin{figure}
\begin{center}
\includegraphics[width=8cm]{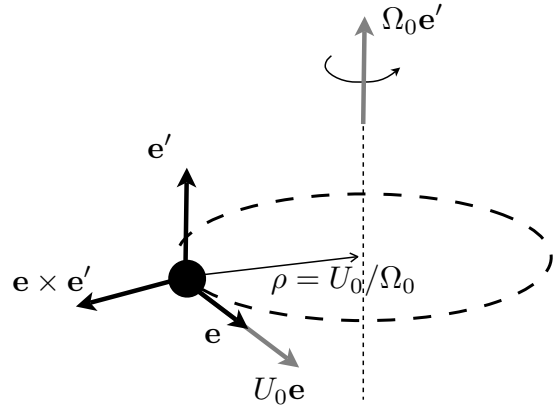}
\caption{Isolated rotating swimmer in a circular trajectory. The intrinsic velocity of the swimmer is the superposition of a translation parallel to $\eb$ and a rotation along $\eb'$. Here, it is assumed that $\eb\cdot\eb'=0$ which leads to a circular trajectory of radius $\rho=U_0/\Omega_0$. The local basis $(\eb,\eb',\eb\times\eb')$ moves rigidly with the swimmer. }\label{fig:figswim1}
\end{center}
\end{figure}

We first consider an isolated swimmer, whose intrinsic motion is the superposition of a translation,  $U_0\eb$, and a rotation, $\Omega_0 \eb'$, where $\eb$ and $\eb'$ are two directions rigidly attached to the swimmer. We neglect here the shape changes of the swimmer, assuming the swimming motion is generated by surface displacements that are small compared to the general dimensions of the swimmer. This is the case for example for a so-called squirmer with tangential displacements for which the shape is at all time a sphere of constant radius \citep{ishikawa06,ishikawa_pedley_rheology07,ishikawa_pedley_diffusion07,ishikawa08}. The two directions $\eb$ and $\eb'$ are fixed in the frame attached to the swimmer and their relative orientation is independent of time. In the absence of Brownian motion, the resulting equations of motion  for the model cell are given by
\begin{equation}
\totd{\rb}{t}=U_0\eb,\quad \totd{\eb}{t}=\Omega_0\eb'\times\eb,\quad\totd{\eb'}{t}=0
\end{equation}
Considering only the non-trivial case where $U_0\neq 0$, three situations can be considered:
\begin{itemize}
\renewcommand{\labelitemi}{$-$}
\item{$\Omega_0=0$: If isolated, the swimmer keeps a fixed orientation and swims along a straight line at constant speed}
\item{$\Omega_0\neq 0$ and $\eb\cdot\eb'=0$: The isolated swimmer has a periodic motion along a circular trajectory of radius $U_0/\Omega_0$ normal to $\eb'$ and the period of the motion is $2\pi/\Omega_0$.}
\item{General case: When $\Omega_0\neq0$ and $\eb\cdot\eb'\neq 0$, the swimmer trajectory is an helix (right-handed if $\eb\cdot\eb'>0$, left-handed otherwise). The pitch of the helix is $(2\pi U_0/\Omega_0)\, \eb\cdot\eb'$, the radius of the circular projection is $U_0/\Omega_0\sqrt{1-(\eb\cdot\eb')^2}$}
\end{itemize}
In this paper, we consider the case of swimmers with circular trajectory, so that $\eb\cdot\eb'=0$ (see illustration in Fig.~\ref{fig:figswim1}).

\subsection{Far-field velocity and vorticity field created by a general swimmer}
In this work, we propose a study of hydrodynamic interactions in the far-field limit, considering only the dominant contribution to the velocity field setup by the self-propelled bodies. The advantage of such an approach is to avoid having to focus on one particular geometry and gait of the swimmer considered. More detailed studies of hydrodynamic interactions can be obtained by considering the full flow field created by a biologically realistic self-propelled cell  \citep{ishikawa06,ishikawa_paramecia06,ishikawa07_bacteria} or for simplified swimmer models \citep{pooley07_2,gyria2009}. In the former case, the flow field must in general be solved for numerically, while in the latter, the simplification of the  geometry and swimming stroke allows for analytical treatment.

In general, a self-propelled cell creates its intrinsic swimming velocity, $\Ub$, and angular velocity, $\boldsymbol\Omega$, by imposing a displacement of its surface. This is the case for all well-studied motile cells, including spermatozoa, bacteria, ciliates and algae. We denote the swimmer surface  $\Sc$. 
This stroke velocity field is noted $\ub^\Sc(\ssb)$ with $\ssb$ the position vector as measured from a point within or in the vicinity of the body position, and fixed in the absolute reference frame. The absolute velocity at the boundary of the swimmer can therefore be written
\begin{equation}
\ub(\ssb)=\Ub+\boldsymbol\Omega\times\ssb+\ub^\Sc(\ssb), \qquad \textrm{for   } \ssb\in\Sc.
\end{equation}
Let $\ub(\xb)$ be the velocity field resulting from this swimming pattern and $\boldsymbol\sigma$ the associated stress field, so that $\fb=\boldsymbol\sigma\cdot\nb$ is  the force per unit area applied by the fluid on the swimmer's boundary with $\nb$ the normal unit vector pointing {into the fluid domain}. The fluid velocity field $\ub$ at a point $\xb$ outside the swimmer can be expressed using the single-layer and double-layer potentials \citep{pozrikidis1997}
\begin{equation}\label{singdoubpot}
u_j(\xb)=-\frac{1}{8\pi\mu}\int_\Sc f_i(\ssb)G_{ij}(\xb,\ssb)\dd S(\ssb)-\frac{1}{8\pi}\int_\Sc u^\Sc_i(\ssb)T_{ijk}(\xb,\ssb)n_k(\ssb)\dd S(\ssb),
\end{equation}
where $G_{ij}(\xb,\ssb)$ is the Green's function corresponding to the flow field at $\xb$ generated by a unit point force located in $\ssb$, and $T_{ijk}(\xb,\ssb)$ is the corresponding stress tensor, and where we have used Einstein's summation notation in  Eq.~\eqref{singdoubpot}. The tensors $G_{ij}$ and $T_{ijk}$ are the Green's function and corresponding stress tensor for the free flow case,
\begin{equation}\label{green}
G_{ij}(\xb,\ssb)=\frac{\delta_{ij}}{r}+\frac{r_ir_j}{r^3},\quad T_{ijk}(\ssb,\xb)=-\frac{6r_ir_jr_k}{r^5},\qquad \textrm{with   }\rb=\xb-\ssb.
\end{equation}
In the far-field approximation, $|\xb|\gg|\ssb|$, and expanding Eq.~\eqref{singdoubpot} in Taylor series, the flow field is obtained as
\begin{eqnarray}\label{singdoubpot2}
u_j(\xb)&=&-\frac{G_{ij}(\xb,0)}{8\pi\mu}\int_\Sc f_i(\ssb)\dd S(\ssb)-\frac{1}{8\pi\mu}\pard{G_{ij}}{s_k}(\xb,0)\int_\Sc f_i(\ssb)s_k\dd S(\ssb)\\ 
&&-\frac{T_{ijk}(\xb,0)}{8\pi}\int_\Sc u^\Sc_i(\ssb)n_k(\ssb)\dd S(\ssb)+O\left(\frac{a^3}{r^3}\right)\nonumber,
\end{eqnarray}
with $a$ the typical size of the swimmer. From Eq.~\eqref{green}, we get
\begin{equation}\label{green2}
\pard{G_{ij}}{s_k}=-\pard{G_{ij}}{r_k}=\frac{\delta_{ij}r_k-\delta_{ik}r_j-\delta_{jk}r_i}{r^3}+\frac{3r_ir_jr_k}{r^5}\cdot
\end{equation}
When $\Rey=0$, the inertia of the swimmer is negligible and the total force and torque 
applied by the fluid on the swimmer must vanish; therefore
\begin{equation}
\int_\Sc f_i(\ssb)\dd S(\ssb)=0,
\end{equation}
and $\int_\Sc f_i(\ssb)s_k\dd S(\ssb)$ must be a symmetric tensor. Consequently,  the first term in Eq.~\eqref{singdoubpot2} vanishes, and only the symmetric part in $i$ and $k$ of $\partial G_{ij}/\partial s_k$, obtained in Eq.~\eqref{green2}, must be retained in Eq.~\eqref{singdoubpot2}. The second and third terms in Eq.~\eqref{singdoubpot2} behave like $1/r^2$ far from the swimmer: The dominant velocity field far from the swimming body is dipolar, and dominated by a so-called stresslet \citep{batchelor1970}
\begin{equation}\label{velindx}
u_i(\rb)=-\frac{3}{8\pi\mu}\left[\frac{r_jr_kS_{jk}}{r^5}\right]r_i+O\left(\frac{a^3}{r^3}\right),
\end{equation}
with the stresslet tensor $\Sb$ given by
\begin{equation}\label{stressletdef}
S_{ij}=\int_\Sc s_if_j(\mathbf{s})\dd\mathbf{s}-\frac{\delta_{ij}}{3}\int_\Sc s_kf_k(\mathbf{s})\dd\mathbf{s}-\mu\int_\Sc \left[u^\Sc_i(\ssb)n_j(\ssb)+u^\Sc_j(\ssb)n_i(\ssb)\right]\dd S(\ssb).
\end{equation}
Note that the definition of the stresslet obtained using the single and double layer potentials is the same as the one obtained by \cite{batchelor1970}.
In the following, we will refer to two different kinds of swimmers, pushers and pullers, by analogy to a simple case where the swimmer can be replaced by a drag-generating center and a thrust-generating center. In that case, all the components of $p_{ij}=\int_\Sc s_if_j\dd S$ are zero except $p_{11}$. For a pusher, the thrust generating center (e.g. flagellum) is located behind the drag-generating center (e.g. head) and $p_{11}<0$. A puller has the opposite configuration and $p_{11}>0$ \citep{laugapowers} (note that $\bf f$ was defined as the force density from the fluid acting on the swimmer, so a pusher acts with a force distribution on the surrounding fluid as directed away from its body along the swimming direction, whereas a puller acts on the fluid with a force distribution directed toward the body along the swimming direction). Finally, by taking the curl of Eq.~\eqref{velindx}, it is straigtforward to get that the vorticity field created by the swimmer is
\begin{equation}\label{vortindx}
\omega_i(\rb)=-\frac{3\epsilon_{ijk}}{4\pi\mu}\frac{r_kr_nS_{jn}}{r^5}+O\left(\frac{a^4}{r^4}\right).
\end{equation}

\def\t{{\rm tr}}

In this work, we will keep the stresslet tensor $\bf S$  general, to model arbitrary swimming modes. Its only constraints are: (1)  ${\bf S}^T={\bf S}$ in order to enforce torque-free motion, and (2) $\t (\bf S)=0$, to ensure the conservation of mass through any closed surfaced enclosing the swimmer.  In general, $\Sb$ depends on the orientation of the swimmer. In the following, we assume that in a frame geometrically attached to the swimmer, the stresslet is time-independent in intensity (eigenvalues of the tensor) and direction (eigenvectors) so that $\Sb=\Rb^T\Sigb\Rb$ where $\Sigb$ is the intrinsic (traceless) stresslet in the set of axes $\mathcal{B}=(\eb,\eb',\eb\times\eb')$ and $\Rb^T$ is the matrix whose columns are the coordinates of $\mathcal{B}$ in the absolute reference frame $\mathcal{B}_0$. As the swimmer moves, $\Rb$(t) depends on time but $\Sigb$ remains constant. Since $\Rb$ is unitary and corresponds to a rotation in three-dimensional space, it corresponds to only {\color{black} three} degrees of freedom. 

In vector notations, Eqs.~(\ref{velindx})-(\ref{vortindx}) become at leading order
\begin{equation}\label{farfieldvel}
\ub(\rb)=-\frac{3}{8\pi\mu}\left[\frac{\rb^T\cdot\Sb\cdot\rb}{r^5}\right]\rb,\quad \omegab=-\frac{3}{4\pi\mu}\frac{(\Sb\cdot\rb)\times\rb}{r^5},\quad {\bf S}^T={\bf S},\quad \textrm{tr}(\Sb)=0.
\end{equation}

\subsection{Coupled motion of two swimmers}
We now consider two identical rotating swimmers, characterized by their position $\rb_j$ and their orientation defined by the two orthogonal vectors $\eb_j$ and $\eb_j'$ ($j=1,2$). The corresponding rotation matrices $\Rb_j$ are defined as above. The problem is non-dimensionalized using the radius $\rho=U_0/\Omega_0$ of the swimmers' circular trajectory and their intrinsic velocity $U_0$. The tensors $\Sb_j$ are scaled using a particular norm $\Lambda$ of $\Sb_j$ (for example the magnitude of its largest eigenvalue) --- $\Lambda$ is an intrinsic property of $\Sigb$ and is therefore identical for both swimmers.

In the far-field approximation, the velocity and rotation of swimmer $1$ induced by swimmer $2$ are respectively equal to the velocity and rotation rate ({\it i.e.} half the vorticity) induced by the motion of swimmer $2$ alone at the position of swimmer $1$. We neglect any higher-order term arising from the finite size of the swimmers  \citep{kimbook}. Such higher order corrections correspond to a modification by the presence of swimmer $1$ of the velocity field created by swimmer $2$. {\color{black} The non-dimensional distance $r$ between the two swimmers must therefore satisfy $r\gg a/\rho$}. To restrict ourselves to the simpler case, we also implicitely assumed that the swimmers are spherical. In the case of a non-spherical swimmer, a correction must be added to the rotation rate even in the far-field approximation, which physically arises from the alignment of an elongated body in a straining (irrotational) flow
\citep{pedley92,laugapowers}. The different limitations introduced by these approximations are discussed in \S\ref{limitations}.

{\color{black} Using the results of the previous sections, and under the assumptions presented above, the dimensionless equations of motion of the coupled swimmers become
\begin{subeqnarray}\label{geneq_each}
\totd{\rb_1}{t}&=&\eb_1-\gamma\left[\frac{(\rb_1-\rb_2)^T\cdot\Sb_2\cdot(\rb_1-\rb_2)}{|\rb_1-\rb_2|^5}\right](\rb_1-\rb_2),\\
\totd{\rb_2}{t}&=&\eb_2-\gamma\left[\frac{(\rb_2-\rb_1)^T\cdot\Sb_1\cdot(\rb_2-\rb_1)}{|\rb_2-\rb_1|^5}\right](\rb_2-\rb_1),\\
\totd{\eb_1}{t}&=&\left\{\eb_1'+\frac{\gamma(\rb_1-\rb_2)\times[\Sb_2\cdot(\rb_1-\rb_2)]}{|\rb_1-\rb_2|^5}\right\}\times\eb_1,\\
\totd{\eb_1'}{t}&=&\left\{\frac{\gamma(\rb_1-\rb_2)\times[\Sb_2\cdot(\rb_1-\rb_2)]}{|\rb_1-\rb_2|^5}\right\}\times\eb_1',\\
\totd{\eb_2}{t}&=&\left\{\eb_2'+\frac{\gamma(\rb_2-\rb_1)\times[\Sb_1\cdot(\rb_2-\rb_1)]}{|\rb_2-\rb_1|^5}\right\}\times\eb_2,\\
\totd{\eb_2'}{t}&=&\left\{\frac{\gamma(\rb_2-\rb_1)\times[\Sb_1\cdot(\rb_2-\rb_1)]}{|\rb_2-\rb_1|^5}\right\}\times\eb_2',
\end{subeqnarray}
where $\gamma=3\Lambda/(8\pi\mu\rho^2 U_0)$. Defining $\rb=\rb_2-\rb_1$, the relative position of the swimmer, and $r=|\rb|$, their relative distance, these equations can be rewritten for the relative motion of the two coupled swimmers as
\begin{subeqnarray}\label{geneq}
\totd{\rb}{t}=\eb_2-\eb_1&-&\gamma\left[\frac{\rb^T\cdot(\Sb_2+\Sb_1)\cdot\rb}{r^5}\right]\rb,\\
\totd{\eb_1}{t}=\left[\eb_1'+\frac{\gamma\rb\times(\Sb_2\cdot\rb)}{r^5}\right]\times\eb_1,&&\totd{\eb_2}{t}=\left[\eb_2'+\frac{\gamma\rb\times(\Sb_1\cdot\rb)}{r^5}\right]\times\eb_2,\\
\totd{\eb_1'}{t}=\left[\frac{\gamma\rb\times(\Sb_2\cdot\rb)}{r^5}\right]\times\eb_1',&&\totd{\eb_2'}{t}=\left[\frac{\gamma\rb\times(\Sb_1\cdot\rb)}{r^5}\right]\times\eb_2',
\end{subeqnarray}
}
Defining $\rb_0=(\rb_1+\rb_2)/2$, the position of the midpoint between the two swimmers, the global motion of the pair of swimmers is given by\begin{equation}\label{geneq_r0}
2\totd{\rb_0}{t}=\eb_2+\eb_1-\left[\frac{\gamma\rb^T\cdot(\Sb_1-\Sb_2)\cdot\rb}{r^5}\right]\rb.
\end{equation}
Eq.~\eqref{geneq} is a system of five vector equations for $\rb$, $\eb_j$ and $\eb'_j$ ($j=1,2$), which is closed
because the knowledge of $\eb_j$ and $\eb_j'$ entirely determines $\Rb_j$ and therefore $\Sb_j$.
Moreover, the equalities  $\eb_j\cdot \eb'_j=0$ and $\eb_j\cdot\eb_j=\eb'_j\cdot\eb'_j=1$ for $j=1,2$ mean that, a priori, Eqs.~\eqref{geneq}-\eqref{geneq_r0} correspond to a twelve-dimensional dynamical system. Eq.~\eqref{geneq} can be solved first for the relative motion since it does not involve $\rb_0$, and one can then obtain the absolute displacement $\rb_0$ from Eq.~\eqref{geneq_r0}.

%% SECTION 3 FAR-FIELD EQUATIONS
\section{Far-field interaction of two rotating swimmers}
\label{multscale}
We are interested in this section in the behavior of Eq.~\eqref{geneq} when the swimmers are far from each other, namely, when their relative distance is much greater than the radius of their trajectory ($r\gg 1$). We can focus our attention to the relative motion of the swimmers defined by $\rb$ as their absolute mean displacement $\rb_0$ does not influence Eqs.~\eqref{geneq}.

Rescaling the distance between the swimmers as $r=r^*/\varepsilon$ with $\varepsilon\ll 1$ and $r^*=O(1)$, the equations for the relative motion are obtained from Eq.~(\ref{geneq}) as (dropping the star superscripts for clarity):
\begin{subeqnarray}\label{scaledeq}
\totd{\rb}{t}&=&\varepsilon(\eb_2-\eb_1)+\varepsilon^3\Fb(\rb,\eb_1,\eb_1',\eb_2,\eb_2'),\\
\totd{\eb_1}{t}&=&\eb_1'\times\eb_1+\varepsilon^3\Gb_1(\rb,\eb_2,\eb_2')\times\eb_1,\quad \totd{\eb_1'}{t}=\varepsilon^3\Gb_1(\rb,\eb_2,\eb_2')\times\eb_1'\slabel{e1scale},\\
\totd{\eb_2}{t}&=&\eb_2'\times\eb_2+\varepsilon^3\Gb_2(\rb,\eb_1,\eb_1')\times\eb_2,\quad \totd{\eb_2'}{t}=\varepsilon^3\Gb_2(\rb,\eb_1,\eb_1')\times\eb_2'\slabel{e2scale},
\end{subeqnarray}
with 
\begin{subeqnarray}
\Fb(\rb,\eb_1,\eb_1',\eb_2,\eb_2')&=&-\gamma\left[\frac{\rb^T\cdot(\Sb_1+\Sb_2)\cdot\rb}{r^5}\right]\rb,\\
\Gb_1(\rb,\eb_2,\eb_2')&=&\frac{\gamma\rb\times(\Sb_2\cdot\rb)}{r^5},\\
\Gb_2(\rb,\eb_1,\eb_1')&=&\frac{\gamma\rb\times(\Sb_1\cdot\rb)}{r^5},
\end{subeqnarray}
which are at most $O(1)$. In addition, differentiating the equations for $\eb_i$ in Eqs.~(\ref{scaledeq}b-c) with respect to time leads to
\begin{equation}\label{deriv2}
\totd{^2\eb_i}{t^2}+\eb_i=\totd{\eb_i'}{t}\times\eb_i+\varepsilon^3\left[\totd{\Gb_i}{t}\times\eb_i+(\Gb_i\cdot\eb_i)\eb_i'-2(\eb_i'\cdot\Gb_i)\eb_i\right]+\varepsilon^6\Gb_i\times(\Gb_i\times\eb_i),
\end{equation}
since $\eb_i\cdot\eb_i'=0$.

\subsection{Multiple-scale analysis}
\label{multscale2}

The equations for $\eb_i$ in Eq.~(\ref{scaledeq}) suggest that in the limit of small $\varepsilon$ there are two different time scales: The short time-scale is $O(1)$ and corresponds to the intrinsic circular motion of the swimmers, whereas the long time scale is $O(\varepsilon^{-3})$ and corresponds to the motion induced on one swimmer by the other. Using the formalism of multiple-scale analysis \citep{benderorszag1978} with the assumption of scale separation arising from the far-field approximation ($\varepsilon\ll 1$), we now formally consider all the fields as functions of two variables $t$ and $\tau=\varepsilon^3 t$. The time derivative operator $\mathrm{d}/\mathrm{d}t$ must then be replaced by $\partial/\partial t+\varepsilon^3\partial/\partial\tau$, and the different vector fields are obtained as regular perturbations series in $\varepsilon$
\begin{subeqnarray}\label{exp}
\rb&=&\rb^{(0)}+\varepsilon\rb^{(1)}+\varepsilon^2\rb^{(2)}+...,\\
\eb_i&=&\eb_i^{(0)}+\varepsilon\eb_i^{(1)}+\varepsilon^2\eb_i^{(2)}+...,\\
\eb_i'&=&\eb_i'^{(0)}+\varepsilon\eb_i'^{(1)}+\varepsilon^2\eb_i'^{(2)}+...,
\end{subeqnarray} 
and the functions $\Fb$, $\Gb_i$ can also be expanded as power series in $\varepsilon$, each term being computed from the expansion of $\eb_i$ and $\rb$. Introducing this expansion in Eq.~(\ref{scaledeq}), we obtain the dynamical system at successive orders, which we now solve. 

At order $O(1)$, we have
\begin{equation}\label{order0}
\pard{\rb^{(0)}}{t}=0,\quad\pard{\eb_i'^{(0)}}{t}=0,\quad\pard{\eb_i^{(0)}}{t}=\eb_i'^{(0)}\times\eb_i^{(0)},
\end{equation}
at order $O(\varepsilon)$
\begin{equation}\label{order1}
\pard{\rb^{(1)}}{t}=\eb_2^{(0)}-\eb_1^{(0)},\quad\pard{\eb_i'^{(1)}}{t}=0,\quad\pard{\eb_i^{(1)}}{t}=\eb_i'^{(0)}\times\eb_i^{(1)}+\eb_i'^{(1)}\times\eb_i^{(0)},
\end{equation}
at order $O(\varepsilon^2)$
\begin{equation}\label{order2}
\pard{\rb^{(2)}}{t}=\eb_2^{(1)}-\eb_1^{(1)},\quad\pard{\eb_i'^{(2)}}{t}=0,\quad\pard{\eb_i^{(2)}}{t}=\eb_i'^{(0)}\times\eb_i^{(2)}+\eb_i'^{(1)}\times\eb_i^{(1)}+\eb_i'^{(2)}\times\eb_i^{(0)},
\end{equation}
and at order $O(\varepsilon^3)$
\begin{eqnarray}\label{order3}
\pard{\rb^{(0)}}{\tau}+\pard{\rb^{(3)}}{t}=\eb_2^{(2)}-\eb_1^{(2)}+\Fb(\rb^{(0)},\eb_1^{(0)},\eb_1'^{(0)},\eb_2^{(0)},\eb_2'^{(0)}),\\
\pard{\eb_i'^{(0)}}{\tau}+\pard{\eb_i'^{(3)}}{t}=\Gb_i(\rb^{(0)},\eb_j^{(0)},\eb_j'^{(0)}),\quad\pard{\eb_i^{(0)}}{\tau}+\pard{\eb_i^{(3)}}{t}=...\nonumber
\end{eqnarray}
Note that we are only interested in the leading order behavior of each function.
Eq.~(\ref{order0}) gives that the leading behavior of $\rb$ and $\eb_i'$ only varies with the long time scale $\tau$. However, it is necessary to go up to the terms of order $O(\varepsilon^3)$ to obtain the $\tau$ dependence of these functions. This results from the ratio between the two time scales being $O(\varepsilon^3)$ while the first correction to $\rb$ is $O(\varepsilon)$. We note from the structure of Eqs.~(\ref{order1})--(\ref{order3}) that the $t$-dependance of the $O(\varepsilon^j)$ term in $\rb$ is determined by the previous order in the expansion of $\eb_i$. We also note that the relation between $\rb^{(j)}$ and $\eb_i^{(j-1)}$ is linear.

If we now introduce the expansion from Eq.~(\ref{exp}) into Eq.~(\ref{deriv2}), we obtain
\begin{equation}
\pard{^2\eb_i^{(j)}}{t^2}+\eb_i^{(j)}=0, \quad 0\leq j\leq 2.
\end{equation}
This equation can be integrated in $t$ as
\begin{equation}\label{eisol}
\eb_i^{(j)}(t,\tau)=\mathbf{a}_i^{(j)}(\tau)\cos(t)+\mathbf{b}_i^{(j)}(\tau)\sin(t).
\end{equation}
If we note $\left\langle.\right\rangle$ the $t$-averaging operator between $t$ and $t+2\pi$, we therefore obtain
\begin{equation}
\left\langle\eb_i^{(j)}\right\rangle(t,\tau)=0,\quad 0\leq j\leq 2,
\end{equation}
and therefore $\left\langle\rb^{(1)}\right\rangle$ and $\left\langle\rb^{(2)}\right\rangle$ are functions of $\tau$ only. We can now take the average of the first equation in Eq.~(\ref{order3}) remembering that $\rb^{(0)}$ is a function of $\tau$ only
\begin{equation}\label{aver}
\totd{\rb^{(0)}}{\tau}-\left\langle\Fb(\rb^{(0)},\eb_1^{(0)},\eb_1'^{(0)},\eb_2^{(0)},\eb_2'^{(0)})\right\rangle=\rb^{(3)}(t,\tau)-\rb^{(3)}(t+2\pi,\tau).
\end{equation}
From Eqs.~(\ref{order0})-(\ref{order2}), $\eb_i'^{(j)}$ is independent of $t$ for $0\leq j\leq 2$. Therefore we have
\begin{equation}\label{eisol2}
\eb_i^{(j)}\times\eb_i'^{(j)}=\mathbf{\tilde{a}}_i^{(j)}(\tau)\cos(t)+\mathbf{\tilde{b}}_i^{(j)}(\tau)\sin(t).
\end{equation}
From the definition of $\Fb$ and Eqs.~(\ref{eisol})-(\ref{eisol2}), we can write
\begin{eqnarray}\label{expF}
\Fb(\rb^{(0)},\eb_1^{(0)},\eb_1'^{(0)},\eb_2^{(0)},\eb_2'^{(0)})&=&\mathbf{A}(\tau)\cos(2t)+\mathbf{B}(\tau)\sin(2t)\\
&&+\mathbf{C}(\tau)\cos(t)+\mathbf{D}(\tau)\sin(t)+\mathbf{E}(\tau),\nonumber
\end{eqnarray}
and the left-hand side of Eq.~(\ref{aver}) is a function $\boldsymbol\alpha(\tau)$ of $\tau$ only. Then we have $\rb^{(3)}(t+2n\pi,\tau)=\rb^{(3)}(t,\tau)-n\alpha(\tau)$. For the perturbation expansion assumption to remain valid, $\alpha$ must be equal to zero. Therefore,
\begin{equation}\label{soldist}
\totd{\rb^{(0)}}{\tau}=\left\langle\Fb(\rb^{(0)},\eb_1^{(0)},\eb_1'^{(0)},\eb_2^{(0)},\eb_2'^{(0)})\right\rangle.
\end{equation}
The same procedure applied to the second equation in Eq.~(\ref{order3}) gives
\begin{equation}\label{aver2}
\eb_i'^{(3)}(t+2\pi,\tau)-\eb_i'^{(3)}(t,\tau)=-\totd{\eb_i'^{(0)}}{\tau}+\left\langle\Gb_i(\rb^{(0)},\eb_j^{(0)},\eb_j'^{(0)})\right\rangle.
\end{equation}
From the definition of $\Gb_i$, $\Gb_i(\rb^{(0)},\eb_j^{(0)})$ can be written in a similar form as $\Fb(\rb^{(0)},\eb_1^{(0)},\eb_1'^{(0)},\eb_2^{(0)},\eb_2'^{(0)})$ in Eq.~(\ref{expF}). The right hand side of Eq.~(\ref{aver2}) is therefore a function of $\tau$ only and to avoid secular terms, both sides of the equation must be zero and
\begin{equation}\label{solorient}
\totd{\eb_i'^{(0)}}{\tau}=\left\langle\Gb_i(\rb^{(0)},\eb_j^{(0)},\eb_j'^{(0)})\right\rangle.
\end{equation}
At leading order, the system behaves therefore as
\begin{equation}
\rb=\rb^{(0)}(\tau)+O(\varepsilon),\quad \eb_i=\eb_i^{(0)}(t,\tau)+O(\varepsilon),\quad \eb_i'=\eb_i'^{(0)}(\tau)+O(\varepsilon),
\end{equation}
with
\begin{subeqnarray}
\totd{\rb^{(0)}}{\tau}=\left\langle\Fb(\rb^{(0)},\eb_1^{(0)},\eb_1'^{(0)},\eb_2^{(0)},\eb_2'^{(0)})\right\rangle&,&\quad\totd{\eb_i'^{(0)}}{\tau}=\left\langle\Gb_i(\rb^{(0)},\eb_j^{(0)},\eb_j'^{(0)})\right\rangle\times\eb_i'^{(0)},\\
\pard{\eb_i^{(0)}}{t}=\eb_i'^{(0)}(\tau)\times\eb_i^{(0)}&,&\pard{\rb^{(1)}}{t}=\eb_2^{(0)}-\eb_1^{(0)}
\end{subeqnarray}

\begin{figure}
\begin{center}
\includegraphics[width=15cm]{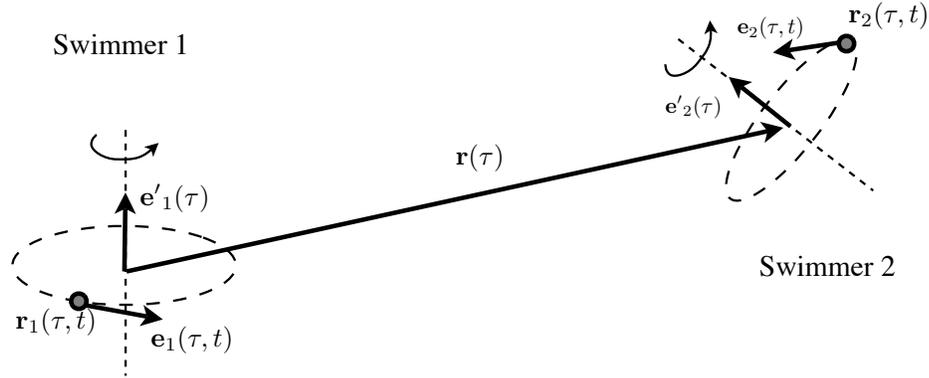}
\caption{Multiple-scale analysis for the motion of the two swimming cells: The leading order motion is characterized by the distance between the mean positions of the two swimmers $\rb$ and the orientation of their rotation vectors $\eb_1'$ and $\eb_2'$. These three vectors evolve with the slow time scale $\tau$, while the instantaneous position of each swimmer is the superposition of their mean and relative motion on the slow time scale $\tau$ and the circular motion on the fast time scale $t$.}\label{fig:figswim2}
\end{center}
\end{figure}

The different notations are summarized on Fig.~\ref{fig:figswim2}, where the superscript $(0)$ was dropped for clarity. We note that to achieve our final result, the hypothesis $\eb_i\cdot\eb_i'=0$ was crucial: It is only because the intrinsic motion produces no net displacement over a period that the separation of scales is possible. If it is not the case but the dot product of these vectors is small, the intrinsic trajectory would be an helix but the net displacement $h$ over one period would still be small. We expect that the analysis remain valid provided $h\ll r$, but this should be confirmed with a perturbation expansion in the helix step, which gives a new small parameter.

\subsection{Computation of the average quantities}
\label{average_equations}
\begin{figure}
\begin{center}
\includegraphics[width=10cm]{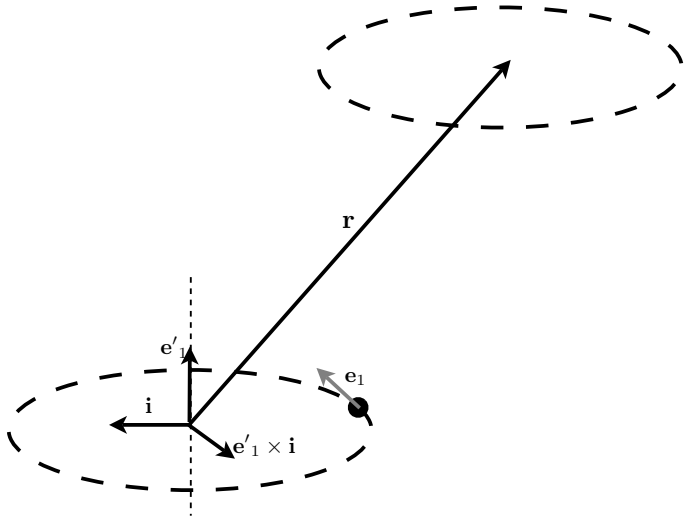}
\caption{Notations for the computation of the average quantities $\left\langle\Fb(\rb,\eb_1,\eb_1',\eb_2,\eb_2')\right\rangle$ and $\left\langle\Gb_i(\rb,\eb_j,\eb_j')\right\rangle$ over a period of the short time scale $t$ corresponding to one period of the circular motion of swimmer $1$. The vectors in black are constant over this time-scale (they depend on $\tau$) and the grey vector $\eb_1$ evolves as Eq.~(\ref{averaging1}).}\label{fig:averagenot}
\end{center}
\end{figure}
In this section, we compute quantities such as $\left\langle\Fb(\rb^{(0)},\eb_1^{(0)},\eb_1'^{(0)},\eb_2^{(0)},\eb_2'^{(0)})\right\rangle$ and $\left\langle\Gb_i(\rb^{(0)},\eb_j^{(0)},\eb_j'^{(0)})\right\rangle$ with the average taken over one period of the short time-scale $t$. For clarity of notations, we drop the $(0)$ exponents with the understanding that we are only considering vector fields of that order. Over this period, $\rb$ and $\eb_j'$ are constant vectors. Defining a unit vector $\ib$ orthogonal to $\eb_i'$ and $\rb$, the basis $\mathcal{B}_p=(\ib,\eb_i'\times\ib,\eb_i')$ is orthonormal (Fig.~\ref{fig:averagenot}). {\color{black} The instantaneous intrinsic directions corresponding to the intrinsic translation and rotation velocities vary as}
\begin{equation}\label{averaging1}
\eb_i=\cos t\,\ib+\sin t\,\eb_i'\times\ib, \qquad \eb_i'\times\eb_i=-\sin t\,\ib+\cos t\,\eb_i'\times\ib
\end{equation}
with no loss of generality since we can redefine the origin of time so that $\eb_i$ is orthogonal to $\rb$ at $t=0$ ($\rb\cdot\ib=0$). The vector $\rb$ can also be decomposed in $\mathcal{B}_p$
\begin{equation}
\rb=r_1\eb_i'+r_2\eb_i'\times\ib\quad \textrm{with   } r_1^2+r_2^2=r^2.
\end{equation}
Then we easily  obtain
\begin{equation}
\Rb\cdot\rb=\left(\begin{array}{c}
\eb_i\cdot\rb\\ \eb_i'\cdot\rb\\ (\eb_i\times\eb_i')\cdot\rb
\end{array}\right)=
\left(\begin{array}{c}
r_2\sin t\\ r_1\\ -r_2\cos t
\end{array}\right),
\end{equation}
and
\begin{eqnarray*}
\left\langle(\eb_i\cdot\rb)^2\right\rangle=\frac{r_2^2}{2},\quad \left\langle(\eb_i'\cdot\rb)^2\right\rangle=r_1^2,\quad\left\langle((\eb_i'\times\eb_i)\cdot\rb)^2\right\rangle=\frac{r_2^2}{2},\\
\left\langle(\eb_i\cdot\rb)(\eb_i'\cdot\rb)\right\rangle=\left\langle(\eb_i\cdot\rb)[(\eb_i'\times\eb_i)\cdot\rb]\right\rangle=\left\langle[(\eb_i'\times\eb_i)\cdot\rb](\eb_i'\cdot\rb)\right\rangle=0,
\end{eqnarray*}
and therefore
\begin{eqnarray}
\left\langle\rb^T\cdot\Sb_i\cdot\rb\right\rangle&=&\left\langle(\Rb\cdot\rb)^T\cdot\Sigb\cdot(\Rb\cdot\rb)\right\rangle\\
\nonumber &=&\frac{r_2^2}{2}\left(\Sigma_{11}+\Sigma_{33}\right)+r_1^2\Sigma_{22}\\
\nonumber &=&\frac{[r^2-(\eb_i\cdot\rb)^2]\mathrm{tr}(\Sigb)}{2}+\Sigma_{22}\left(\frac{3}{2}(\eb_i'\cdot\rb)^2-\frac{r^2}{2}\right).
\end{eqnarray}
Finally, since $\textrm{tr}(\Sigb)=0$,
\begin{equation}\label{Fav}
\left\langle\Fb(\rb^{(0)},\eb_1^{(0)},\eb_1'^{(0)},\eb_2^{(0)},\eb_2'^{(0)})\right\rangle=\frac{\gamma\Sigma_{22}}{2}\left[\frac{2r^2-3\left[(\eb_1'\cdot\rb)^2+(\eb_2'\cdot\rb)^2\right]}{r^5}\right]\rb.
\end{equation}
Similarly,
\begin{eqnarray}
\Sb_i\cdot\rb&=&\left[\Sigma_{11}r_2\sin t+\Sigma_{12}r_1-\Sigma_{13}r_2\cos t\right]\eb_i\\
\nonumber &&+\left[\Sigma_{21}r_2\sin t+\Sigma_{22}r_1-\Sigma_{23}r_2\cos t\right]\eb_i'\\
\nonumber  &&+\left[\Sigma_{31}r_2\sin t+\Sigma_{32}r_1-\Sigma_{33}r_2\cos t\right](\eb_i\times\eb_i'),
\end{eqnarray}
and 
\begin{eqnarray}
\eb_i\times\rb&=&r_1\sin t\,\ib-r_1\cos t\,(\eb_i'\times\ib)+r_2\cos t\,\eb_i',\\
\eb_i'\times\rb&=&-r_2\ib,\\
 (\eb_i\times\eb_i')\times\rb&=&-r_1\cos t\,\ib-r_1\sin t\,(\eb_i'\times \ib)+r_2\sin t\,\eb_i'
\end{eqnarray}
from which we obtain after time-averaging,
\begin{equation}
\left\langle(\Sb_i\cdot\rb)\times\rb\right\rangle=\frac{r_1r_2}{2}\left(\Sigma_{11}-2\Sigma_{22}+\Sigma_{33}\right)\ib+\left(\Sigma_{31}-\Sigma_{13}\right)\left[\frac{r_2^2}{2}\eb_i'-\frac{r_1r_2}{2}\eb_i'\times\ib\right].
\end{equation}
The last term in the last equation is equal to zero as $\Sigb$ is symmetric, and identifying $r_1=\eb_i'\cdot\rb$ and $r_2\ib=\rb\times\eb_i'$, the previous equation becomes:
\begin{equation}
\left\langle(\Sb_i\cdot\rb)\times\rb\right\rangle=-\frac{3\Sigma_{22}}{2}(\eb_i'\cdot\rb)(\rb\times\eb_i').
\end{equation}
Therefore,
\begin{eqnarray}
\left\langle\Gb_1(\rb^{(0)},\eb_2^{(0)},\eb_2'^{(0)})\right\rangle&=&\frac{3\gamma\Sigma_{22}(\eb_2'\cdot\rb)(\rb\times\eb_2')}{2r^5},\\
 \left\langle\Gb_2(\rb^{(0)},\eb_1^{(0)},\eb_1'^{(0)})\right\rangle&=&\frac{3\gamma\Sigma_{22}(\eb_1'\cdot\rb)(\rb\times\eb_1')}{2r^5}.\nonumber
\end{eqnarray}
Finally, the relative equations of motion for the slow varying fields $\rb$, $\eb_1'$ and $\eb_2'$ become with $\mu=\gamma\Sigma_{22}$
\begin{subeqnarray}\label{sloweq}
\totd{\rb}{\tau}&=&\mu\left[\frac{2r^2-3\left[(\eb_1'\cdot\rb)^2+(\eb_2'\cdot\rb)^2\right]}{2r^5}\right]\rb\\ 
\totd{\eb_1'}{\tau}&=&\frac{3\mu(\eb_2'\cdot\rb)[(\rb\times\eb_2')\times\eb_1']}{2r^5},\\
\totd{\eb_2'}{\tau}&=&\frac{3\mu(\eb_1'\cdot\rb)[(\rb\times\eb_1')\times\eb_2']}{2r^5}.
\end{subeqnarray}
The absolute motion of the swimmers can be determined on the long time scale $\tau$ by averaging \eqref{geneq_r0} over the short-time scale and obtain
\begin{equation}\label{average_abs}
\totd{\langle\rb_0\rangle}{\tau}=\frac{3\mu}{4}\left[(\eb'_2\cdot\rb)^2-(\eb'_1\cdot\rb)^2\right]\frac{\rb}{r^3}.
\end{equation}
A comparison of the dynamical systems given by  Eq.~(\ref{sloweq}) and Eq.~(\ref{geneq}) shows that the averaged equations, Eq.~(\ref{sloweq}), correspond to the interaction of two stresslets of equal intensity $3\mu/2(\eb_1'\eb_1'-\mathbf{I}/3)$ and $3\mu/2(\eb_2'\eb_2'-\mathbf{I}/3)$ respectively located at the mean position of swimmers $1$ and $2$ with no intrinsic velocity. This suggests that a single swimmer creates an average far-field in the form of a stresslet whose intensity is $3\mu/2$ and whose orientation is entirely determined by its intrinsic rotation vector $\eb_i'$. This statement is proven rigorously in \S\ref{average_stresslet}. The intensity of the averaged stresslet is equal to $3\Sigma_{22}/2$, where $\Sigma_{22}$ is the diagonal component of the instantaneous stresslet along the direction $\eb_i'$. We observe that all the other components of $\Sigb$ disappear in the averaging process.

By analogy with the case where the instantaneous stresslet is equal to a force dipole, resulting from the superposition of a drag force and a thrust force, we will consider in the following two kinds of swimmers:
\begin{itemize}
\item{\textit{Pushers} with $\mu>0$: In this case the thrust generating center is located behind the drag generating center; $\gamma\Sigma_{11}<0$ and $\gamma\Sigma_{22}=\gamma\Sigma_{33}>0$ with all other components equal to zero [see Eq.~\eqref{stressletdef}]. This is for example the case of a swimmer with a flagellum located behind its drag-generating head, such as spermatozoa, or most flagellated bacteria.}
\item{\textit{Pullers} with $\mu<0$: In that case, the thrust is generated in front of the drag-generating center; $\gamma\Sigma_{11}>0$ and $\gamma\Sigma_{22}=\gamma\Sigma_{33}<0$ [see Eq.~\eqref{stressletdef}]. This is for example the case for swimmers using their flagella in a breaststroke pattern to pull their bodies, such as the alga {\it Chlamydomonas}.}
\end{itemize}

It is important to point out here that we manage to obtain a system of equations for $\eb_1'$, $\eb_2'$ and $\rb$ only, but that the position of each swimmer on its instantaneous circular trajectory is not important --- in particular the relative phase of these instantaneous motions. {\color{black} Two conditions are necessary for this simplification to occur. First, the average flow field created by an isolated rotating swimmer is independent of time and also independent of the direction of motion on the circular trajectory (see the following section). This is a consequence of the fact that the instantaneous flow field created by the swimmer does not have any azimuthal component. The second condition is that the swimmers are spherical, and the averaged velocity induced on swimmer $2$ by swimmer $1$ only depends on the properties of the averaged flow field induced by swimmer $1$ and not the orientation of swimmer $2$. This would not be the case if the swimmers were non-spherical: then, the induced velocity and rotation created by swimmer $1$ on swimmer $2$ would not only depend on the position and trajectory of swimmer $1$, but also on the orientation of swimmer $2$ with respect to the principal axes of strain of the local flow (see the discussion in \S\ref{limitations}). For non-spherical swimmers, the averaging process is more subtle and the phase of the instantaneous motions of the two swimmers does not disappear in the averaged equations; it remains however a constant parameter of the problem since both swimmers have the same intrinsic translation and rotation velocities.}

\subsection{Far-field averaged velocity field created by a rotating swimmer}
\label{average_stresslet}
The results of the previous section suggest that, on average, a rotating swimmer behaves like a stresslet in the far-field. We explore this result in more detail in this section. The behavior of the far-field velocity is of interest to characterize the rheological properties of a suspension of such swimmers \citep{batchelor1970}. In this section only, we consider an isolated swimmer, and compute the time-averaged flow in the far field. The swimmer trajectory is a circle oriented by its rotation vector $\eb'$ parallel to the vertical axis and we choose the origin of the reference axes as the average position of this swimmer. Let denote by $\epsb(t)$ the instantaneous position of the swimmer ($|\epsb(t)|=1$ by our choice of scaling) and $\eb$ its velocity vector. If $\ib$ is an arbitrary constant unit vector orthogonal to $\eb'$, we can define the origin of time such that:
\begin{equation}\label{smmotion}
\epsb(t)=\cos t\,\ib+\sin t\,\eb'\times\ib,\quad \eb=-\sin t\,\ib+\cos t\,\eb'\times \ib.
\end{equation}
We are interested in the velocity field created by this swimmer at a position $\xb$ far from the origin ($x\gg 1$). The instantaneous velocity field at $\xb$ is given from Eq.~(\ref{farfieldvel}) by 
\begin{equation}
\ub(\xb)=-\gamma\left[\frac{(\Rb\cdot\rb)^T\cdot\Sigb\cdot(\Rb\cdot\rb)}{r^5}\right]\rb,\textrm{   with   }\rb=\xb-\epsb,
\end{equation}
and 
\begin{equation}
\mathbf{P}=\Rb\cdot\rb=\left(\begin{array}{c} \eb\cdot\rb\\\eb'\cdot\rb\\(\eb\times\eb')\cdot\rb\end{array}\right)=\left(\begin{array}{c} \eb\cdot\xb\\\eb'\cdot\xb\\(\eb\times\eb')\cdot\xb\end{array}\right)-\left(\begin{array}{c} \eb\cdot\epsb\\\eb'\cdot\epsb\\(\eb\times\eb')\cdot\epsb\end{array}\right)=\left(\begin{array}{c} \eb\cdot\xb\\\eb'\cdot\xb\\(\eb\times\eb')\cdot\xb\end{array}\right)-\left(\begin{array}{c} 0\\0\\1\end{array}\right).
\end{equation}
Therefore from Eq.~(\ref{smmotion}), noting once again $\left\langle.\right\rangle$ the averaging operator over a $2\pi$-period, we have
\begin{subeqnarray}
\left\langle P_1^2\right\rangle&=&\frac{1}{2}\left[\left(\xb\cdot(\eb'\times\ib)\right)^2+\left(\ib\cdot\xb\right)^2\right]=\frac{1}{2}\left[x^2-\left(\xb\cdot\eb'\right)^2\right],\\
\left\langle P_2^2\right\rangle&=&\left(\xb\cdot\eb'\right)^2,\\
\left\langle P_3^2\right\rangle&=&1+\frac{1}{2}\left[x^2-\left(\xb\cdot\eb'\right)^2\right],\\
\left\langle P_2P_3\right\rangle&=&-\xb\cdot\eb',\\
\left\langle P_1P_2\right\rangle&=&\left\langle P_1P_3\right\rangle=0.
\end{subeqnarray}
Keeping only the dominant terms, we have on average
\begin{equation}
\left\langle \rb^T\cdot\Sb\cdot\rb\right\rangle=\frac{\Sigma_{11}+\Sigma_{33}}{2}\left[x^2-\left(\xb\cdot\eb'\right)^2\right]+\Sigma_{22}\left(\xb\cdot\eb'\right)^2=\frac{\Sigma_{22}}{2}\left[3(\xb\cdot\eb')^2-x^2\right].
\end{equation}
We also have
\begin{equation}
\frac{1}{r^n}=\frac{1}{x^n}\left(1+n\frac{\epsb\cdot\xb}{x^2}+o\left(\frac{1}{x}\right)\right).
\end{equation}
Since we are interested only in the dominant term in the far-field averaged behavior, we write
\begin{equation}
\left\langle\frac{(\Rb\cdot\rb)^T\cdot\Sigb\cdot(\Rb\cdot\rb)}{r^5}\right\rangle\sim\frac{\left\langle(\Rb\cdot\rb)^T\cdot\Sigb.(\Rb\cdot\rb)\right\rangle}{x^5},
\end{equation}
as all the corrections to this expression are of higher order in $1/x$. Grouping all terms, we finally obtain the far-field averaged flow
\begin{equation}\label{farfieldav}
\left\langle\ub\right\rangle(\xb)=-\frac{\gamma\Sigma_{22}}{2}\frac{\left[\xb^T\cdot\left(3\eb'\eb'-\mathbf{I}\right)\cdot\xb\right]\xb}{x^3}\cdot
\end{equation}
We recognize here the velocity field created by a steady stresslet $3\mu/2\left(\eb'\eb'-\mathbf{I}/3\right)$ consistently with the results of the previous section. Physically, the results of Eq.~\eqref{farfieldav} indicate that, for cells which behave instantaneously as pushers (pullers), the averaged flow is that of a puller (pusher) along the axis of rotation of the circular motion.

{\color{black} We observe in Eq.~\eqref{farfieldav} that the average flow remains identical by changing $\eb'$ into $-\eb'$: the average flow is therefore not modified by a reversal of the circular motion (along the same trajectory).}

%% SECTION 4 DYNAMICAL SYSTEM ANALYSIS

\section{Analysis of the far-field interaction}
\label{sec:dynsys}
\subsection{Reduced forms of the equations}
\label{redeq}

\begin{figure}
\begin{center}
\includegraphics[width=10cm]{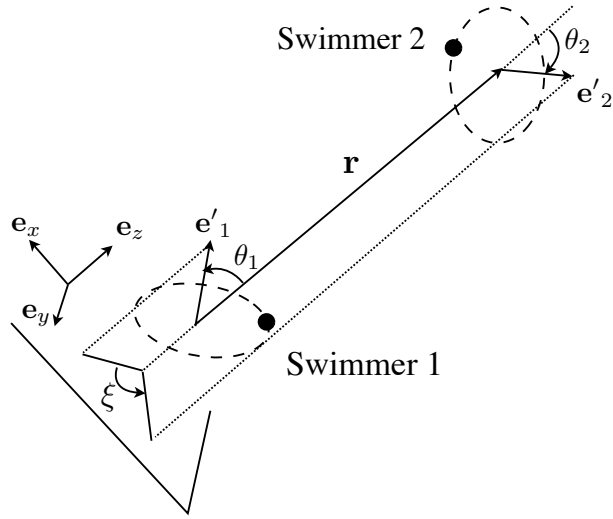}
\caption{Definitions of the various variables for the average motion (see text for details).}\label{fig:figswim4}
\end{center}
\end{figure}

We now return to the coupled equations derived using the multiple-scale analysis. Defining the unit vector $\eb_z$ of the direction between swimmer $1$ and swimmer $2$, $\eb_z=\rb/|\rb|$ and by differentiation in time we obtain
\begin{equation}\label{direction}
\totd{\eb_z}{\tau}=\frac{1}{|\rb|}\totd{\rb}{\tau}-\left(\frac{\rb}{|\rb|^3}\cdot\totd{\rb}{\tau}\right)\rb.
\end{equation}
But from Eq.~(\ref{sloweq}), we note that $\mathrm{d}\rb/\mathrm{d}\tau=\mathcal{R}\,\rb$, with $\mathcal{R}$ a scalar function of $\rb$ and $\eb_j'$. Using this result in \eqref{direction}, we obtain that $\eb_z=\rb/|\rb|$ is a time-independent unit vector set by the initial conditions. The mean distance between the two swimmers maintain a fixed direction. In the following, $\eb_z$ denotes {the fixed direction between the two swimmers' positions}. The vectors $\eb_i'$ are defined from $\eb_z$ by their polar and azimuthal angle $\theta_i$ and $\phi_i$. Choosing two constant unit vectors $\eb_x$ and $\eb_y$ so that ($\eb_x$,$\eb_y$,$\eb_z$) is orthonormal, then
\begin{equation}
\eb_i'=\sin\theta_i\cos\phi_i\eb_x+\sin\theta_i\sin\phi_i\eb_y+\cos\theta_i \eb_z.
\end{equation}
Note here, that the definition of $\phi_i$ depends on the definition of $\eb_x$ and $\eb_y$ which can be rotated arbitrarily in the plane orthogonal to $\eb_z$. Therefore, only the intrinsic $\xi=\phi_2-\phi_1$ has a physical meaning. Then in the frame ($\eb_x$,$\eb_y$,$\eb_z$) we have
\begin{equation}
(\eb_1'\cdot\rb)[\rb\times\eb_1']\times\eb_2'=r^2\left(\begin{array}{c}\cos\theta_1\cos\theta_2\sin\theta_1\cos\phi_1\\
\cos\theta_1\cos\theta_2\sin\theta_1\sin\phi_1\\
-\cos\theta_1\sin\theta_1\sin\theta_2\cos(\phi_2-\phi_1)
\end{array}\right),
\end{equation}
and
\begin{equation}
\totd{\eb_2}{\tau}=\totd{\theta_2}{\tau}\left(\begin{array}{c}\cos\theta_2\cos\phi_2\\\cos\theta_2\sin\phi_2\\-\sin\theta_2
\end{array}\right)+\totd{\phi_2}{\tau}\left(\begin{array}{c} -\sin\theta_2\sin\phi_2\\\sin\theta_2\cos\phi_2\\0
\end{array}\right),
\end{equation}
By identification, the system given by Eq.~(\ref{sloweq}) can then be rewritten as a four-dimensional dynamical system
\begin{subeqnarray}\label{sloweqsimp}
\totd{r}{\tau}&=&\frac{\mu}{2r^2}\left[2-3(\cos^2\theta_1+\cos^2\theta_2)\right]\slabel{sloweqsimp1},\\
\totd{\theta_1}{\tau}&=&\frac{3\mu}{2r^3}\cos\theta_2\sin\theta_2\cos\xi,\\
\totd{\theta_2}{\tau}&=&\frac{3\mu}{2r^3}\cos\theta_1\sin\theta_1\cos\xi,\\
\sin\theta_1\sin\theta_2\totd{\xi}{\tau}&=&-\frac{3\mu}{2r^3}\cos\theta_1\cos\theta_2(\sin^2\theta_1+\sin^2\theta_2)\sin\xi,
\end{subeqnarray}
where we have used $\xi=\phi_2-\phi_1$. The notations for Eq.~\eqref{sloweqsimp} are summarized on Fig.~\ref{fig:figswim4}.
Note that Eq.~(\ref{sloweqsimp}) can be simplified even further by defining $\alpha=2r^3/3\mu$, $x_i=\cos\theta_i$ and $y=\sin\theta_1\sin\theta_2\cos\xi$, and we obtain
\begin{subeqnarray}\label{eqred}
\totd{\alpha}{\tau}&=&2-3(x_1^2+x_2^2)\slabel{eqred1},\\
\alpha\totd{x_1}{\tau}&=&-x_2y,\\
\alpha\totd{x_2}{\tau}&=&-x_1y\slabel{eqred2},\\
\alpha\totd{y}{\tau}&=&x_1x_2(2-x_1^2-x_2^2).\slabel{eqred3}
\end{subeqnarray}
Physically, $\alpha$ is proportional to the third power of the distance between the swimmers. It is negative for $\mu<0$ (pullers) and positive for $\mu>0$ (pushers). From the original physical problem, we also have the following three mathematical constraints:
\begin{itemize}
\item{The variable $\alpha$ is either positive or negative. A change of sign of $\alpha$ requires a cancellation of $r$ at a finite time and a collision of the swimmers. Such a collision obviously violates the far-field approximation, and the present theory is not valid when $\alpha$ gets small. In the following, we will refer as ``collisions'' to regimes where the present theory predicts a decrease of the relative distance to an arbitrary small number, at which point additional modeling is required. We will therefore focus on solutions for which the sign of $\alpha$ is fixed.}
\item{The variables $x_1$ and $x_2$ are cosines, therefore $-1\leq\{ x_1,x_2\}\leq 1$.}
\item{From the definition of $y$, $0\leq y^2\leq(1-x_1^2)(1-x_2^2)$.}
\end{itemize}

\subsection{Relative equilibria and stability}
\label{equilibrium}
We focus here on relative equilibrium positions, for which on the long time scale, the swimmers do not move relatively to each other. There can be however a mutual motion of the swimmers ($\mathrm{d}\langle\rb_0\rangle/\mathrm{d}\tau\neq0$).
\subsubsection{Equilibrium points}
From \eqref{eqred}, there is only one type of equilibrium points obtained for $(\alpha,x_1,x_2,y)=(\alpha_0,\pm\sqrt{2/3},0,0)$, or symmetrically $(\alpha,x_1,x_2,y)=(\alpha_0,0,\pm\sqrt{2/3},0)$ for any value  $\alpha_0$ of $\alpha$. 

Physically, the distance between the swimmers can take an arbitrary value but the orientations of the rotation vectors must correspond to a very specific configuration. One swimmer's rotation axis makes an angle $\cos^{-1}\sqrt{2/3}\approx 35\,^\circ$ with the distance between the swimmers. The second swimmer's rotation axis is orthogonal to the first swimmer's and their relative distance (so $\eb'_j$ is orthogonal to the plane defined by $\eb'_i$ and $\eb_z$, with $j\neq i$). 

The linearized system about one such equilibrium is obtained as:
\begin{equation}
\totd{}{t}\left(\begin{array}{c} \alpha-\alpha_0 \\ x_1-\sqrt{\frac{2}{3}}\\ x_2\\ y\end{array}\right)=\left(\begin{array}{cccc} 0 &-\sqrt{6}& 0& 0\\ 0& 0& 0& 0\\ 0& 0& 0& -\frac{1}{\alpha_0}\sqrt{\frac{2}{3}}\\ 0 &0 &\frac{4}{3\alpha_0}\sqrt{\frac{2}{3}} &0\end{array}\right).\left(\begin{array}{c} \alpha-\alpha_0 \\ x_1-\sqrt{\frac{2}{3}}\\ x_2\\ y\end{array}\right)
\end{equation}

The eigenvalues of the above matrix are $\lambda=\pm\ci 2\sqrt{2}/3\alpha_0$, and $\lambda=0$ with multiplicity $2$. The dimension of the subspace associated with $\lambda=0$ is however equal to $1$. It is therefore not possible to conclude from the linearized system on the stability of the equilibrium of the non-linear system as one of the eigenvalue of the linearized system is identically zero (neutral stability) \citep{sastry1999}. We will show rigorously in \S \ref{regimes} that this equilibrium is unstable.

\subsubsection{Rotational equilibria}
Another situation of interest is the case where the direction of the circular motions, $\eb'_i$ remains fixed relatively to $\eb_z$. Only $\alpha$ (or equivalently the distance between the two swimmers) depends on time. This occurs for two different configurations.

\paragraph{Swimmers with same axis of rotation: $x_1=\pm 1$ and $x_2=\pm 1$.}

The two swimmers have quasi-circular trajectories in two parallel planes and $\eb_i'$ are both aligned with $\eb_z$. As a direct consequence of the definitions of $x_i$ and $y$, $y$ must be zero at all time. From Eq.~(\ref{eqred1}), the evolution of $\alpha$  in that configuration can be computed
\begin{equation}\label{coax}
\alpha=\bar\alpha=\alpha_0-4\tau,\quad \textrm{and  }r=\bar{r}=\left(r^3_0-6\mu \tau\right)^{1/3}.
\end{equation}
The overbar denotes the reference configuration (rotational equilibrium) we are considering. Therefore, $\bar{x}_1=\pm 1$ and $\bar{x}_2=\pm 1$. Swimmers with $\mu<0$ (pullers) tend to repel each other while swimmers with $\mu>0$ (pushers) attract each other, until the scale-separation assumptions of the multiple-scale analysis break down. {\color{black} We observe that the collision time scales like $r_0^3/\mu\sim R^3/\rho a l$ with $R$ the dimensional distance between the swimmers, $\rho$ the radius of their circular trajectory and $l$ and $a$ the length and head radius of the swimmer, respectively. }

The stability of this time-varying configuration is now investigated by decomposing each variable $f$ (with $f=\alpha,x_1, x_2,y$) as $f=\bar{f}+f'$ and $f'$ is a small perturbation. At leading order, Eq.~(\ref{eqred}) can be rewritten
\begin{subeqnarray}
\totd{\alpha'}{\tau}&=&-6(\bar{x}_1x_1'+\bar{x}_2x_2')\\
\bar{\alpha}\totd{x_1'}{\tau}=-\bar{x}_2y'&,&\bar{\alpha}\totd{x_2'}{\tau}=-\bar{x}_1y'\\
\alpha\totd{y'}{\tau}&=&-2\bar{x}_1\bar{x}_2(\bar{x}_1x_1'-\bar{x}_2x_2')
\end{subeqnarray}
or equivalently
\begin{equation}
\totd{\alpha'}{\tau}=-6(\bar{x}_1x_1'+\bar{x}_2x_2'),\qquad
\bar\alpha\totd{}{t}\left(\begin{array}{c} x_1'\\x_2'\\y'\end{array}\right)=\left(\begin{array}{cccc}0 & 0& -\bar{x}_2\\0 & 0& -\bar{x}_1\\-2\bar{x}_2& -2\bar{x}_1& 0  \end{array}\right)\cdot\left(\begin{array}{c}  x_1'\\ x_2'\\ y'\end{array}\right)
\end{equation}
where the bar quantities correspond to the relative equilibrium ($\bar\alpha=\alpha_0-4\tau$ and $\bar{x}_i^2=1$) and the prime quantities are perturbations.

The last system can be solved exactly if diagonalized. Defining
\begin{equation}
\left(\begin{array}{c}z_1\\z_2\\z_3\end{array}\right)=\left(\begin{array}{ccc}\bar{x}_2& -\bar{x}_1 &\bar{x}_1\\ \bar{x}_1& -\bar{x}_2&\bar{x}_2\\0& 2&2\end{array}\right)\cdot\left(\begin{array}{c}x_1'\\x_2'\\y'\end{array}\right),\qquad \left(\begin{array}{c}x_1'\\x_2'\\y'\end{array}\right)=\frac{1}{4}\left(\begin{array}{ccc}2\bar{x}_2 & -2\bar{x}_1 & 0\\ \bar{x}_1 & \bar{x}_2 & 1\\ -\bar{x}_1& -\bar{x}_2 & 1\end{array}\right)\cdot\left(\begin{array}{c}z_1\\z_2\\z_3\end{array}\right),
\end{equation}
it decomposes into
\begin{equation}
\totd{z_1}{\tau}=0,\quad \totd{z_2}{\tau}=-\frac{2z_2}{\alpha_0-4\tau},\quad\totd{z_3}{\tau}=\frac{2z_3}{\alpha_0-4\tau}
\end{equation}
which can be integrated easily into
\begin{equation}
z_1=z_{1,0},\quad z_2=z_{2,0}\left(1-\frac{4\tau}{\alpha_0}\right)^{1/2},\quad z_3=z_{3,0}\left(1-\frac{4\tau}{\alpha_0}\right)^{-1/2}.
\end{equation}
The original variables are obtained by linear combinations of these solutions, and we observe that the configuration is unstable with algebraically growing perturbations.

\begin{figure}
\begin{center}
\includegraphics[width=16cm]{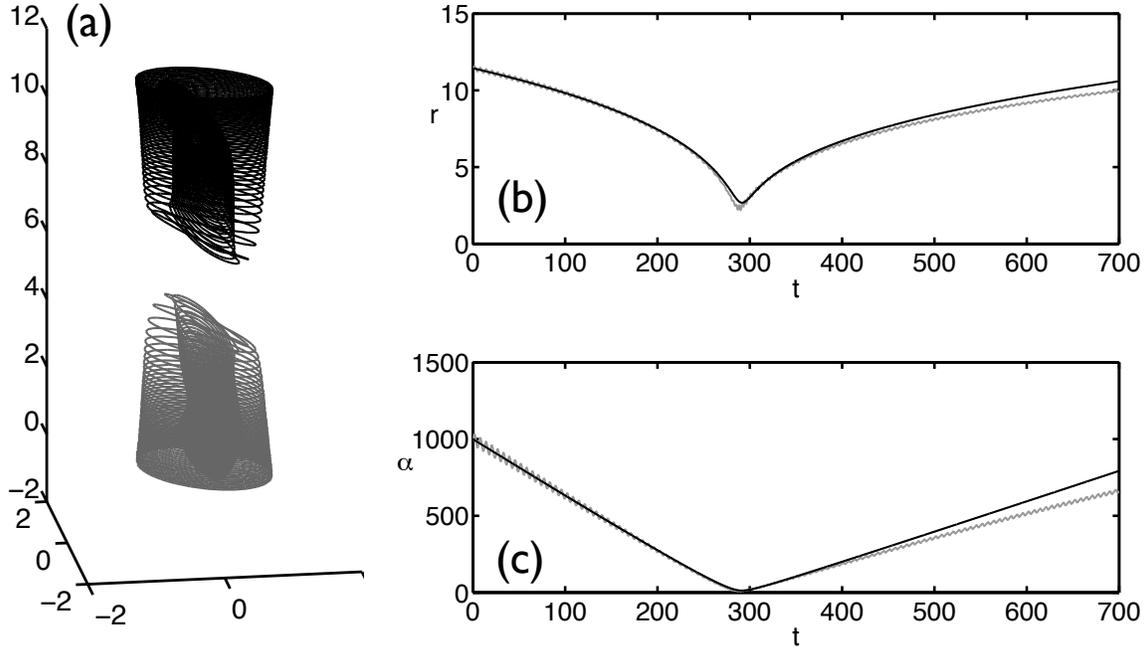}
\caption{{\color{black}Interaction of two pushers ($\mu=1$) with circular motions that are initially almost coaxial ($\theta_1=\theta_2=0.22$, $\xi=0.68$) and an initial distance equal to $r=11.4$. (a) Trajectories of the two swimmers: the initially coaxial pushers attract each other until hydrodynamic interactions modify the orientation of their circular trajectories and they become co-planar, leading to a repulsive interaction. (b) Evolution of the distance between the swimmers and (c) evolution of the parameter $\alpha$. In both (b) and (c), the light grey curve corresponds to the full equations (Eq.~\ref{geneq}) for which the circular motion of each swimmer is resolved, and the black curve corresponds to the simplified model for the averaged motion (Eq.~\ref{eqred}). Note that the two curves agree with each other until the swimmers get close to each other.}}\label{coaxcoplan}
\end{center}
\end{figure}

{\color{black}
Figure \ref{coaxcoplan} illustrates this situation and compares the prediction of the far-field model for the averaged motion (Eq.~\ref{eqred}) to the full set of equations (Eq.~\ref{geneq}). Considering two pushers ($\mu>0$) that have initially almost the same axis of rotation ($\theta_1,\theta_2\ll 1$), the hydrodynamic interactions create a mutual attraction of the swimmers following the approximate law (Eq.~\ref{coax}). This rotational equilibrium is unstable and as they get closer from each other, the planes of the trajectories of the two swimmers undergo a quick rotation, bringing the two swimmers from a co-axial to a co-planar configuration (see next section) in which the interaction of the two pushers have now a repulsive effect. Figure \ref{coaxcoplan} also allows to show the agreement between the simplified model (Eq.~\ref{eqred}) and the full equations of the system.
}

\paragraph{Two-dimensional configuration: $x_1=x_2=0$ and $y=\pm 1$.}

For $\theta_i=\pi/2$ and $\xi=0$ ($y=1$) for co-rotating and $\xi=\pi$ ($y=-1$) for counter-rotating swimmers, both swimmers are in the same plane with their rotation axes orthogonal to the plane of motion. Note that the two-dimensional configurations are actually only particular cases of orientational equilibria: $x_i=0$ and $-1\leq y\leq 1$. As above, this configuration is a rotational equilibrium only, as the distance between the swimmers varies in time according to
\begin{equation}
\alpha=\alpha_0+2\tau,\quad r=\left(r_0^3+3\mu\tau\right)^{1/3}.
\end{equation}
This time, swimmers with $\mu<0$ (pullers) attract each other while swimmers with $\mu>0$ (pushers) repel each other. Here again, a stability analysis can be performed, and the linearized dynamics becomes
\begin{equation}
\totd{\alpha'}{\tau}=-3(x_1^{'2}+x_2^{'2}),\qquad
\bar\alpha\totd{}{t}\left(\begin{array}{c} x_1'\\x_2'\\y'\end{array}\right)=\left(\begin{array}{ccc}0 & -\bar{y}& 0\\-\bar{y}&0 & 0\\0 & 0& 0  \end{array}\right)\cdot\left(\begin{array}{c}  x_1'\\ x_2'\\ y'\end{array}\right),
\end{equation}
which can be integrated exactly as
\begin{equation}
y'=y'_0,\quad \frac{x_1'+x_2'}{2}=\left(\frac{x_1'+x_2'}{2}\right)\left(1+\frac{2\tau}{\alpha_0}\right)^{-\bar{y}/2},\quad \frac{x_1'-x_2'}{2}=\left(\frac{x_1'-x_2'}{2}\right)\left(1+\frac{2\tau}{\alpha_0}\right)^{\bar{y}/2},
\end{equation}
once again leading to the instability of these configurations with algebraically growing perturbations.

\subsection{Reduction to a two-dimensional problem}
\subsubsection{Conserved quantities}
\label{conservation}
The system given by Eq.~\eqref{eqred} can be simplified even further by observing that 
\begin{equation}
A=x_1^2-x_2^2=\cos^2\theta_1-\cos^2\theta_2
\end{equation}
is a conserved quantity. Without any loss of generality, we can assume $A$ to be positive (the equations are symmetric with respect to a switch between $x_1$ and $x_2$). In the $(x_1,x_2)$-plane, the system moves along a hyperbola, and we can define the parametric coordinate $\sigma$ such that 
\begin{equation}\label{chgvarx}
x_1=\sqrt{A}\cosh\sigma,\quad x_2=\sqrt{A}\sinh\sigma.
\end{equation}
To be rigorous, $x_1$ should be equal to $\pm\sqrt{A}\cosh\sigma$. However, one can change $(x_1,x_2,y)$ into $(-x_1,-x_2,y)$ by changing the definition of $\eb_z$ to $-\eb_z$ (or equivalently, switching the indices of the swimmers), and we therefore restrict ourselves to $x_1\geq 0$ by redefining $\sigma$ appropriately. Introducing this change of variables into Eq.~\eqref{eqred2} leads to 
\begin{equation}\label{yeq}
y=-\alpha\dot\sigma.
\end{equation}
Using this relation in Eq.~\eqref{eqred3}, we obtain
\begin{equation}
-\alpha(\dot\alpha\dot\sigma+\alpha\ddot\sigma)=\frac{A}{2}\sinh 2\sigma\left(2-A\cosh 2\sigma\right),
\end{equation}
and multiplying by $\dot\sigma$ and integrating with respect to time, we obtain
\begin{equation}
\alpha^2\dot\sigma^2=-A\cosh 2\sigma +\frac{A^2}{2}\cosh^2 2\sigma +C,
\end{equation}
where $C$ is a constant of integration. Noting from Eq.~\eqref{chgvarx} that $A\cosh 2\sigma=x_1^2+x_2^2$, we have therefore proven that
\begin{equation}\label{ACdef}
A=x_1^2-x_2^2\quad\textrm{and}\quad C=y^2+(x_1^2+x_2^2)\left(1-\frac{x_1^2+x_2^2}{4}\right)
\end{equation}
are two conserved quantities in this problem.
Finally, defining the new variable $X=x_1^2+x_2^2=A\cosh 2\sigma$, the system given by Eq.~\eqref{eqred} is equivalent to 
\begin{subeqnarray}\label{eqredb}
\alpha\dot{X}&=&\epsilon\sqrt{\left(X^2-A^2\right)\left(4C-4X+X^2\right)}\slabel{eqredb1}\\
\dot\alpha&=&2-3X\slabel{eqredb2}
\end{subeqnarray}
with $\epsilon=\pm 1$. We have therefore transformed the four-dimensional system, Eq.~\eqref{eqred}, into a two-dimensional system, Eq.~\eqref{eqredb}. The values of the constants $A$ and $C$, as well as the initial values of $X$ and $\alpha$  can be obtained from the initial conditions of the four variables $(\alpha,\theta_1,\theta_2,\xi)$. The choice of the sign of $\epsilon$ is discussed in \S\ref{sign_e}.

\subsubsection{Bounds on the different variables}
From the constraints detailed at the end of \S\ref{redeq} and the definitions of the variable $X$, and the constants $A$ and $C$, we have the following four constraints.
\begin{itemize}
\item{
{$0\leq \{x_1^2,x_2^2\}\leq 1$ therefore 
\begin{subeqnarray}\label{constrAX}
&0\leq A\leq 1,&\\
&A\leq X\leq 2-A. &
\end{subeqnarray}}
\item{
From Eq.~\eqref{ACdef}, $C=y^2+X-X^2/4$. Using the previous bounds on $X$ as well as the inequality $y^2\leq(1-x_1^2)(1-x_2^2)$, we obtain that 
\begin{equation}\label{Cmax}
C\leq 1-\frac{A^2}{4}\cdot
\end{equation}}}
\item{
From Eq.~\eqref{ACdef}, we have $y^2=C-X+X^2/4$, and therefore 
\begin{equation}\label{Xmax}
X\leq 2\left(1-\sqrt{1-C}\right).
\end{equation}
Because of Eq.~\eqref{Cmax}, we have
\begin{equation}
2-A\geq 2\left(1-\sqrt{1-C}\right),
\end{equation}
and Eq.~\eqref{Xmax} is actually a tigher upper bound than Eq.~(\ref{constrAX}b).}
\item{ Finally, Eq.~\eqref{Xmax} and $X\geq A$ implies that $C\geq A-A^2/4$.}
\end{itemize}

In summary, the following inequalities must be satisfied
\begin{equation}\label{bounds}
0\leq A\leq 1,\qquad A-\frac{A^2}{4}\leq C\leq 1-\frac{A^2}{4},\qquad A\leq X\leq 2\left(1-\sqrt{1-C}\right).
\end{equation}

\subsubsection{Choosing the sign of $\epsilon$}\label{sign_e}
For given values of $A$ and $C$, and given initial conditions on $X$ and $\alpha$, there are two possible solutions corresponding to $\epsilon=\pm 1$ initially. In Eq.~\eqref{eqredb1}, the square-root of the right-hand-side is positive and $\epsilon$ has therefore the sign of $\alpha\dot{X}=-4x_1x_2y$. Differentiating with respect to time and using Eq.~\eqref{eqred}, we obtain
\begin{equation}\label{2ndder}
\totd{}{t}\left(\alpha\dot{X}\right)=\frac{4X}{\alpha}\left(C-X+\frac{X^2}{4}\right)+\frac{(X^2-A^2)(X-2)}{\alpha},
\end{equation}
and $\dot{X}$ and $\ddot{X}$ are continuous functions of time. 

From the constraints of  \S \ref{redeq}, we are only interested in solutions where the sign of $\alpha$ is fixed. Therefore, we are not interested in the solutions of Eq.~\eqref{eqredb} past a zero of $\alpha$. The left-hand side of Eq.~\eqref{eqredb1} vanishes only for vanishing $\dot{X}$ or for collisions. We prove here, that if at $t=t_0$, $\alpha\dot{X}=0$, then $\epsilon$ must change of sign at $t=t_0$ if $\alpha(t_0)\neq 0$. Such a cancellation of the left-hand sign of Eq.~\eqref{eqredb1} happens only in two configurations
\begin{enumerate}
\item{$X=A=X_{min}$ or equivalently $x_2=0$ (one swimmer's rotation axis is orthogonal to the distance between the two swimmers). For $X$ to reach a minimum at $t=t_0$, $\dot{X}<0$ for $t<t_0$ and $\epsilon(t_0^-)=\epsilon^-=-\textrm{sgn}(\alpha(t_0))$. For small $|t-t_0|$, we obtain using Taylor expansion and Eq.~\eqref{2ndder},
\begin{equation}
\alpha\dot{X}\sim\left[\frac{4A}{\alpha}\left(C-A+\frac{A^2}{4}\right)\right](t-t_0),
\end{equation}
which is positive for $t>t_0$, therefore $\epsilon(t_0^+)=\epsilon^+=\textrm{sgn}(\alpha(t_0))=-\epsilon^-$.}
\item{$X=2(1-\sqrt{1-C})=X_{max}$ or equivalently $y=0$. For $t<t_0$, we therefore have $\epsilon^-=\textrm{sgn}(\alpha(t_0))$. For small $|t-t_0|$, we obtain
\begin{equation}
\alpha\dot{X}\sim-\left[\frac{2\sqrt{1-C}\left(4(1-\sqrt{1-C})^2-A^2\right)}{\alpha}\right](t-t_0),
\end{equation}
and $\epsilon^+=-\textrm{sgn}(\alpha(t_0))=-\epsilon^-$.}
\end{enumerate}
With the analysis above, we see that for given values of $A$ and $C$ the system can be represented solely in the $(X,\alpha)$ plane. However, if one wants to look at maps of the flow, two maps should be superimposed $\epsilon=1$ and $\epsilon=-1$, one for trajectories of decreasing $X$ and the other for trajectories of increasing $X$.

\subsection{Possible regimes in the far-field interaction of two rotating swimmers}
\label{regimes}
\subsubsection{Monotonic variations of $\alpha$}
From Eq.~\eqref{eqredb2}, we see that $\alpha$ is an increasing (decreasing) function of time if $X\leq 2/3$ ($X\geq 2/3$). If $2/3$ is out of the bounds imposed on $X$ by Eq.~\eqref{bounds}, $\alpha$ and the distance between the swimmers are monotonic functions of time. Two such cases can occur.
\begin{enumerate}
\item{If $A>2/3$, then $\dot\alpha<2-3A<0$, and
\begin{itemize}
\item{if $\alpha_0<0$, $\alpha\rightarrow -\infty$  and the swimmers get further and further away from each other,}
\item{if instead $\alpha_0>0$, $\alpha\rightarrow 0$ and a collision occur at a finite time (since the time derivative of $\alpha$ is negative and has a non-zero negative upper bound.}
\end{itemize}
}
\item{If $C<5/9$, then $\dot\alpha>2-6(1-\sqrt{1-C})>0$ and
\begin{itemize}
\item{if $\alpha_0<0$, $\alpha\rightarrow 0$ and a collision occurs at finite time (since the time derivative of $\alpha$ is positive and has a non-zero positive lower bound),}
\item{if instead $\alpha_0>0$, $\alpha\rightarrow \infty$ and the distance between the swimmers is unbounded.}
\end{itemize}
}
\end{enumerate}

\subsubsection{General case: Theory}\label{theory}
If $A<2/3$ and $C>5/9$, then we can prove that $X$ oscillates from its lower bound $X_{min}=A$ to its upper bound $X_{max}=2(1-\sqrt{1-C})$. This statement could be proven rigorously from the equations for $X$ and $\alpha$. We only provide here a qualitative argument for clarity. Since $\dot{X}$ only vanishes at these bounds, $X$ varies monotonically from one to the other. If $X$ doesn't reach the next bound (even as $t\rightarrow\infty$), then it would have a finite limit and $\dot{X}$ must go to zero as $t\rightarrow\infty$ while $\alpha\dot{X}$ remains finite; this combination can only occur if $\alpha$ is unbounded. Therefore, $X$ oscillates between its bounds unless  $|\alpha|\rightarrow\infty$.\\

Then, let $t_n$ be the successive times at which $X$ reaches either $X_{min}$ or $X_{max}$ and the corresponding values $\alpha_n$. We are interested in the gain $G_n=|\alpha_{n+1}/\alpha_n|$ and the time interval $\tau_n=t_{n+1}-t_n$ between two sign reversals of $\dot{X}$. From Eq.~\eqref{eqredb}, we obtain that, over an interval where $\dot{X}$ has a given sign, we have
\begin{equation}\label{int}
\mathcal{F}(X;A,C,X_0)=\int_{X_0}^X\frac{(2-3X)\dd X}{\sqrt{(X^2-A^2)(4C-4X+X^2)}}=\epsilon\log\left|\frac{\alpha}{\alpha_0}\right| \cdot
\end{equation}
$\mathcal{F}$ is well defined for $X_{min}\leq X\leq X_{max}$ as the singularities at the end points are integrable.
Using Eq.~\eqref{int} between $t_n$ and $t_{n+1}$, we obtain the following.
\begin{itemize}
\item{If $\alpha_0>0$ (therefore $\alpha>0$ at all time at least until collision), then when $X$ varies from $X_{min}$ to $X_{max}$, $\epsilon$ is positive, and
\begin{equation}
\log G_n=\mathcal{G}(A,C), \textrm{  with  } \mathcal{G}(A,C)=\int_{X_{min}}^{X_{max}}\frac{(2-3X)\dd X}{\sqrt{(X^2-A^2)(4C-4X+X^2)}}\cdot
\end{equation}
}
\item{If $\alpha_0<0$ (therefore $\alpha<0$), then when $X$ varies from $X_{min}$ to $X_{max}$, $\epsilon$ is negative and
\begin{equation}
\log G_n=- \mathcal{G}(A,C).
\end{equation}
}
\end{itemize}
We note here that $G_n$ is a function of $A$ and $C$ only and therefore not a function of $n$ or $\alpha_n$. We can therefore summarize these results for all initial choice of $(\alpha,x_1,x_2,y)$ or equivalently $(\alpha_0,X_0,A,C)$:
\label{4cases}
\begin{enumerate}
\item{If $\alpha_0<0$ (puller) and $\mathcal{G}(A,C)>0$, $\alpha\rightarrow 0$ and there is a collision between the swimmers.}
\item{If $\alpha_0<0$ (puller) and $\mathcal{G}(A,C)<0$, $\alpha\rightarrow -\infty$ and the swimmers move away from each other.}
\item{If $\alpha_0>0$ (pusher) and $\mathcal{G}(A,C)>0$, $\alpha\rightarrow \infty$ and the swimmers move away from each other.}
\item{If $\alpha_0>0$ (pusher) and $\mathcal{G}(A,C)<0$, $\alpha\rightarrow 0$ and there is a collision between the swimmers.}
\end{enumerate}
Note that the particular cases discussed in the previous section ($A>2/3$ and $C<5/9$) are also included in this analysis: the integrand in $\mathcal{G}$ has then a fixed sign. 

\begin{figure}
\begin{center}
\begin{tabular}{cc}
\subfigure[\, Pushers ($\mu>0$)]{\includegraphics[width=7cm]{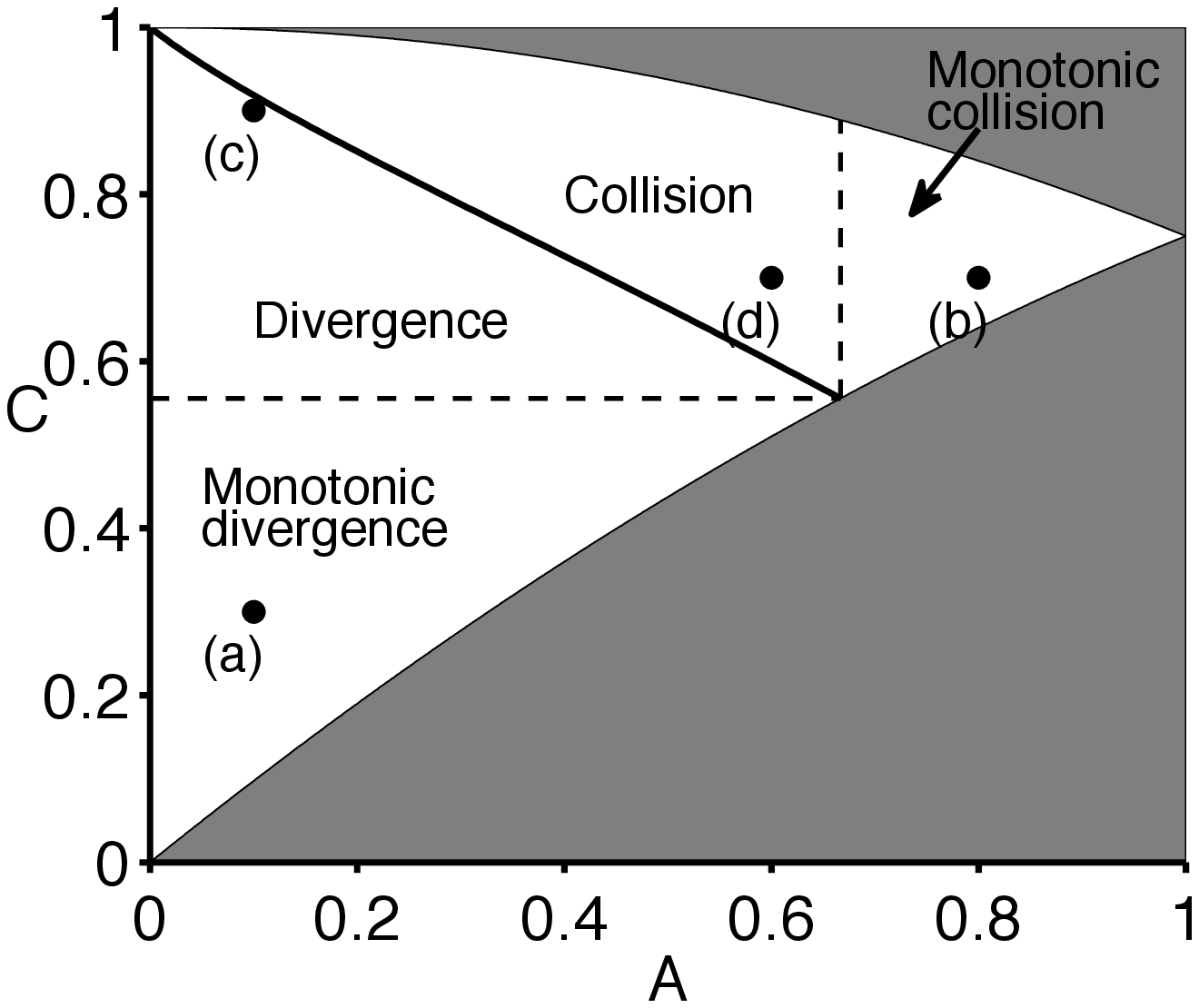}} &
\subfigure[\, Pullers ($\mu<0$)]{\includegraphics[width=7cm]{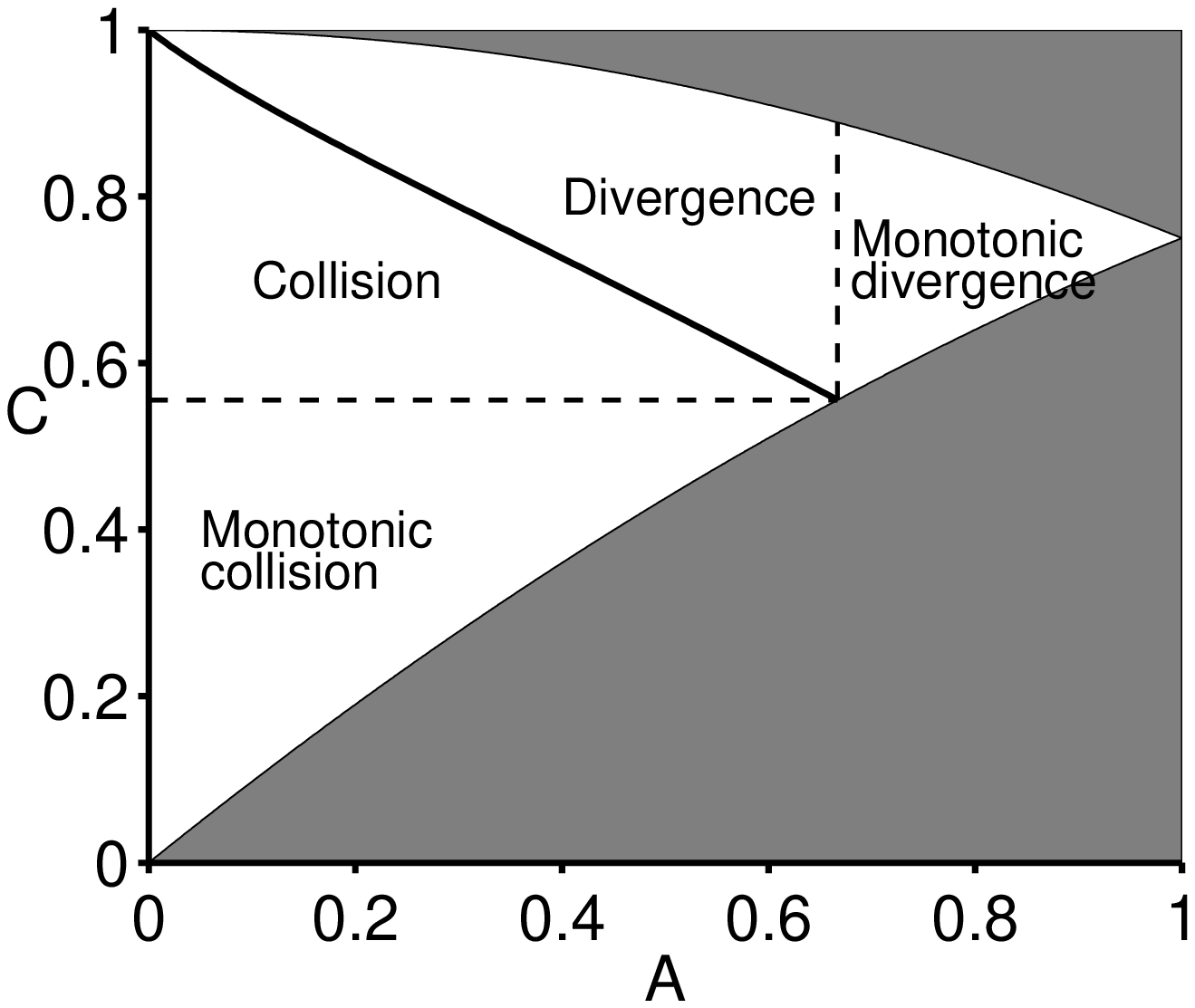}}
\end{tabular}
\caption{Maps of the general regime in the $(A,C)$-plane for the long-time relative behavior of two swimmers with positive $\mu$ (left, pushers) or negative $\mu$ (right, pullers). Note that one can be deduced from the other by symmetry, {\it i.e.} by changing collision (divergence) by divergence (collision). On the left map (pushers), the position of the four examples (a)--(d) of Fig.~\ref{fig:dynsyssol} are indicated.
}\label{fig:regimes}
\end{center}
\end{figure}

It is important to point out that $\mathcal{G}$ determines the regime (divergence or collision) of the two swimmers and is a function of $A$ and $C$ only. The regime is therefore entirely determined by these two quantities, which are only functions of the relative orientation of the rotation vectors of the swimmers and independent of their initial separation distance. The maps of the regimes obtained for pushers ($\mu>0$) and pullers ($\mu<0$) are displayed on Fig.~\ref{fig:regimes}.
Note that at the boundary between the collision and divergence domains, we have $\mathcal{G}=0$: The distance between the swimmers remains unchanged between $t_n$ and $t_{n+2}$ and the motion is periodic. This corresponds to a limit cycle. The boundary between the regimes shown on Fig.~\ref{fig:regimes} (solid line between the divergence and collision regions) can be obtained numerically by finding the values of $A$ and $C$ for which $\mathcal{G}(A,C)=0$. The other two solid boundaries correspond to $C=1-A^2/4$ and $C=A-A^2/4$ [see the constraints on $C$ in \eqref{bounds}].

\subsubsection{General case: Numerical simulations}

\begin{figure}
\begin{center}
\begin{tabular}{ccc}
\includegraphics[width=5cm]{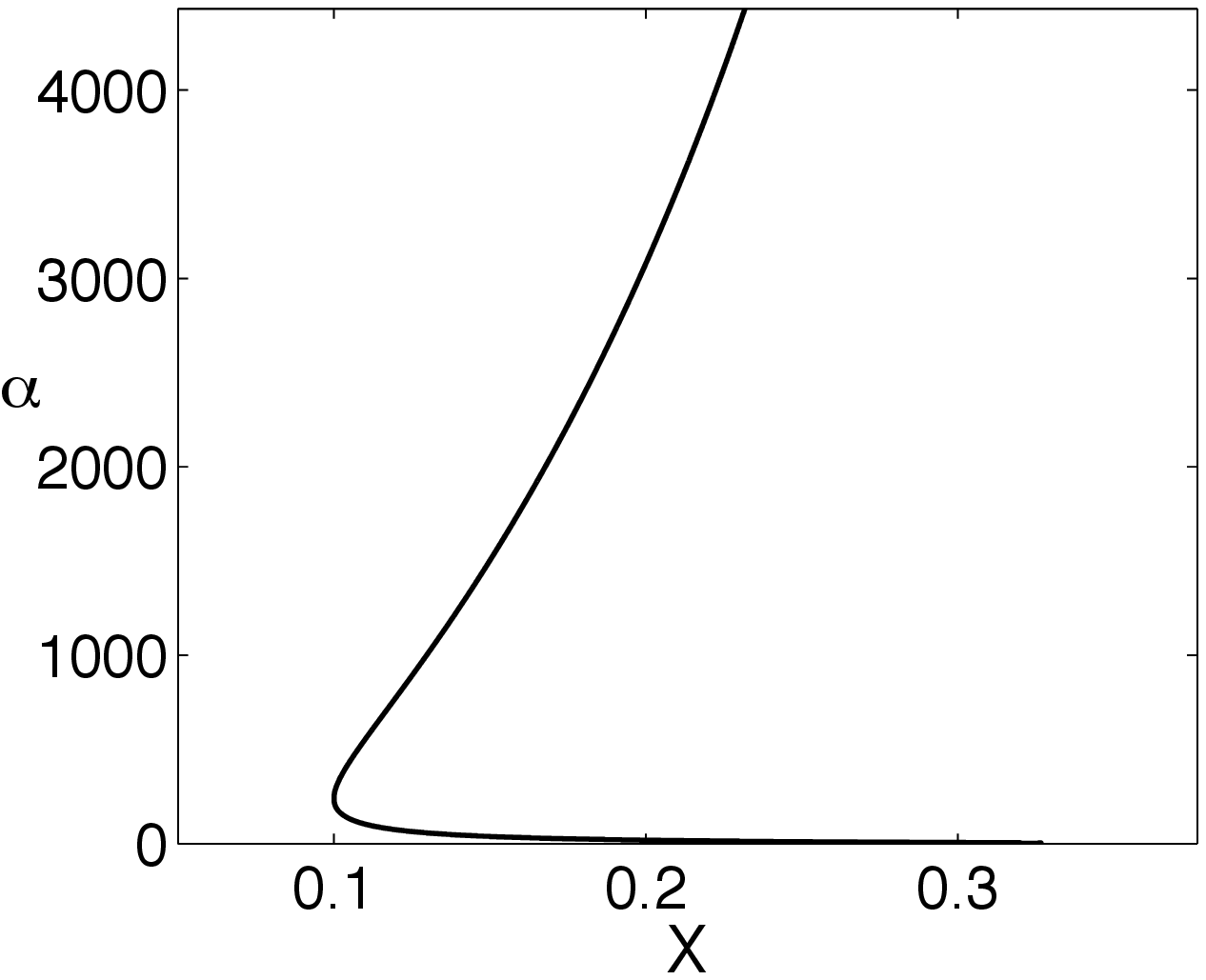} &
\includegraphics[width=5cm]{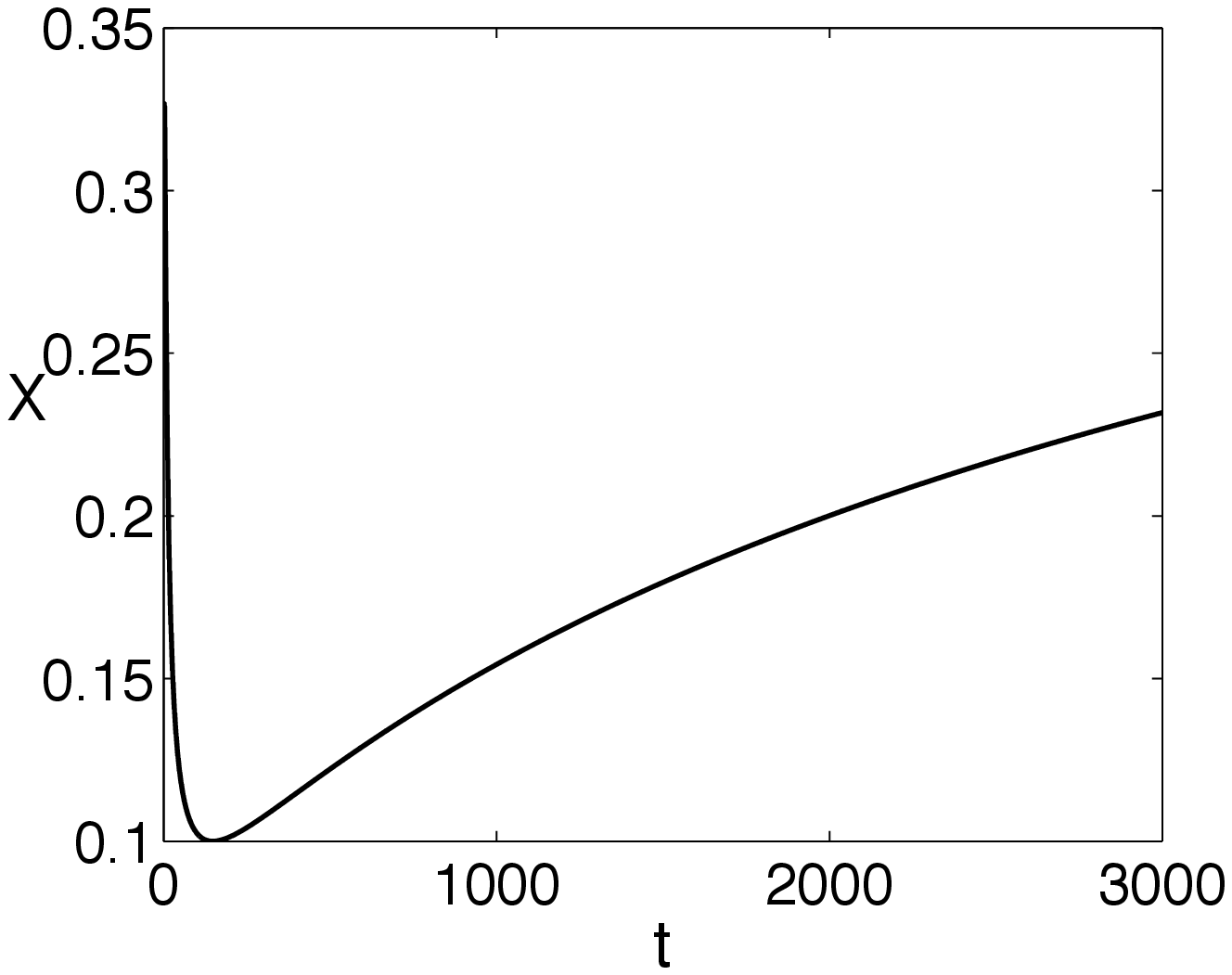} &
\includegraphics[width=5cm]{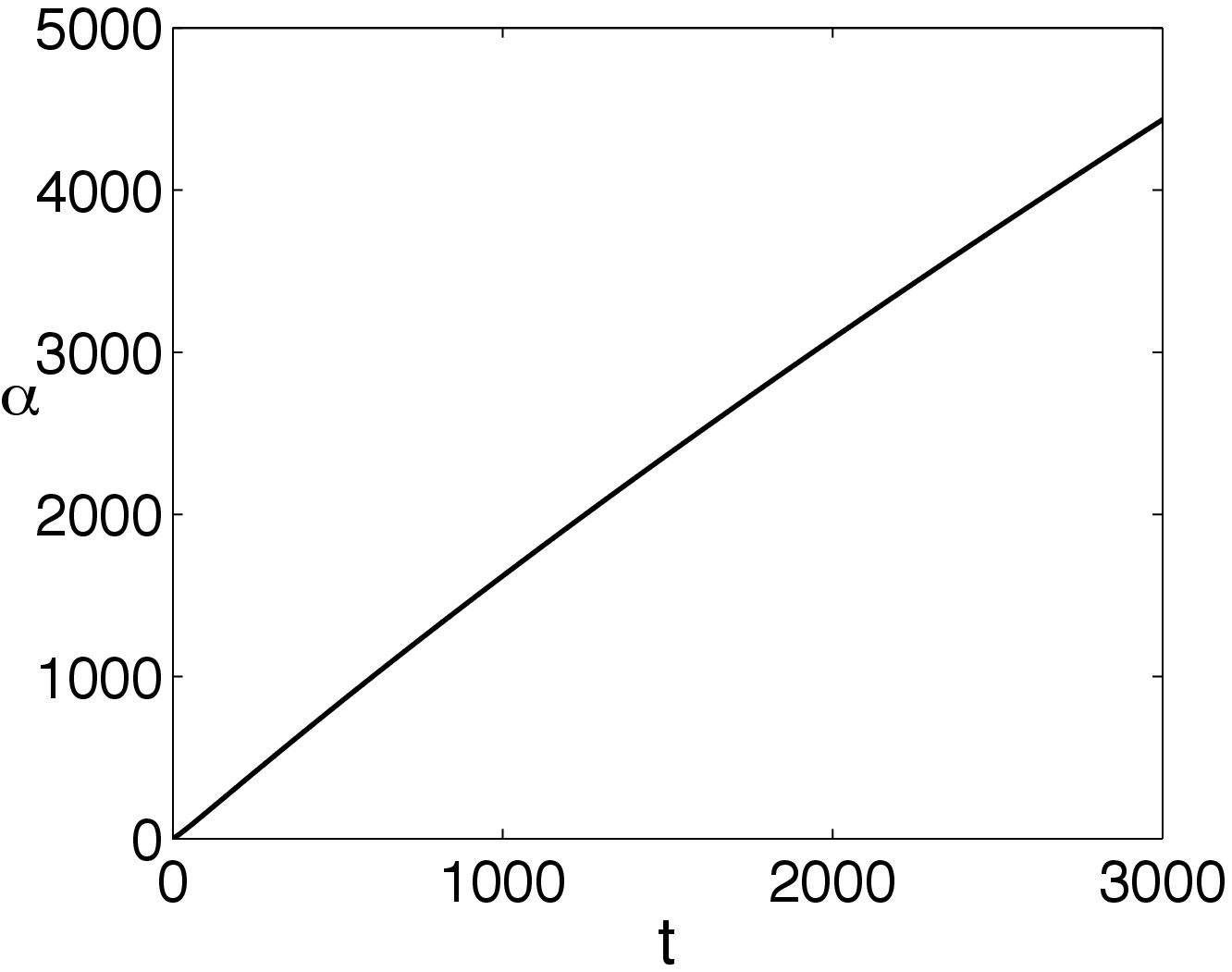} \\
\multicolumn{3}{c}{(a) $A=0.1$, $C=0.3$, $X_0=0.25$ and $\alpha_0=1$}\\
\\
\includegraphics[width=5cm]{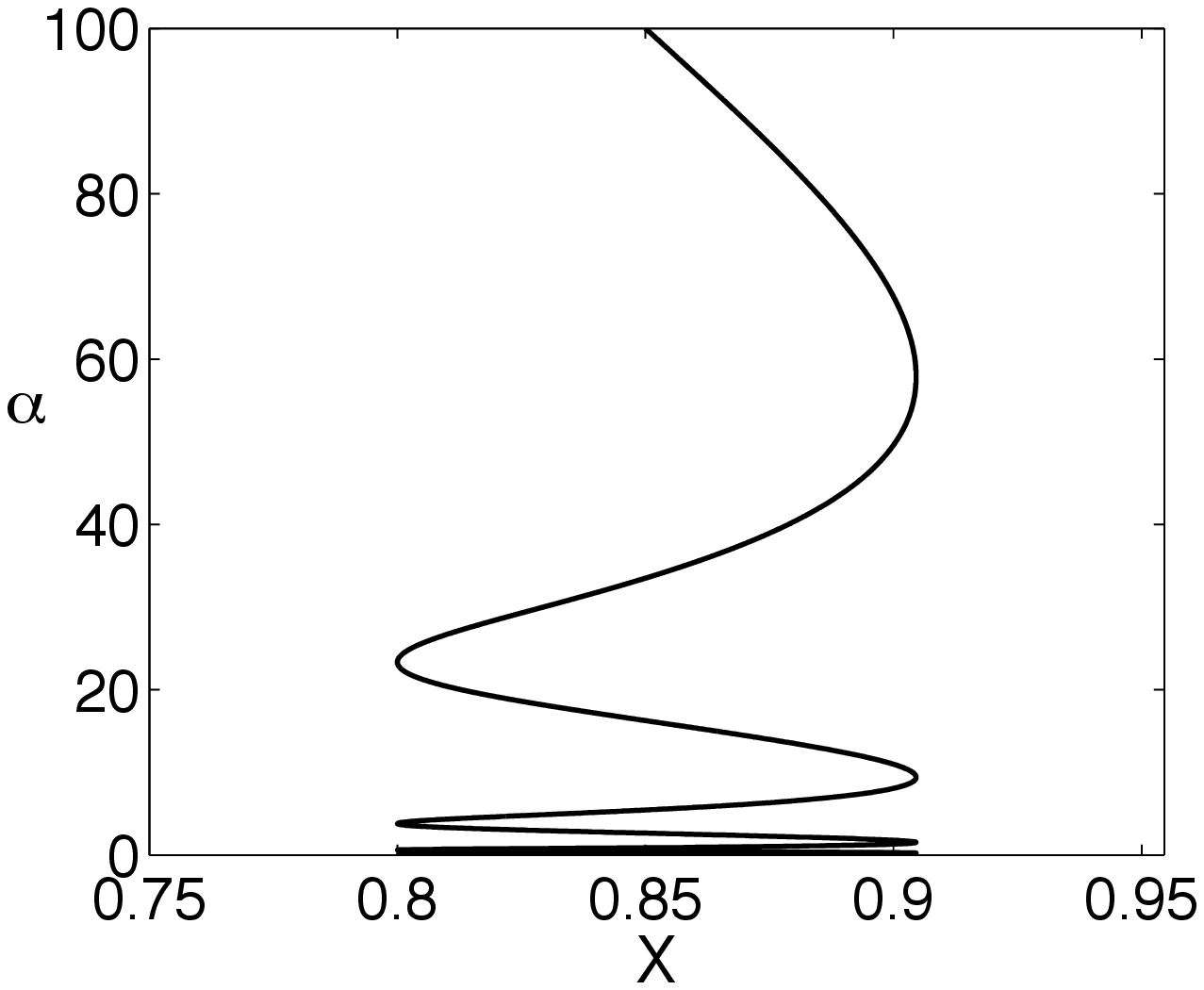} &
\includegraphics[width=5cm]{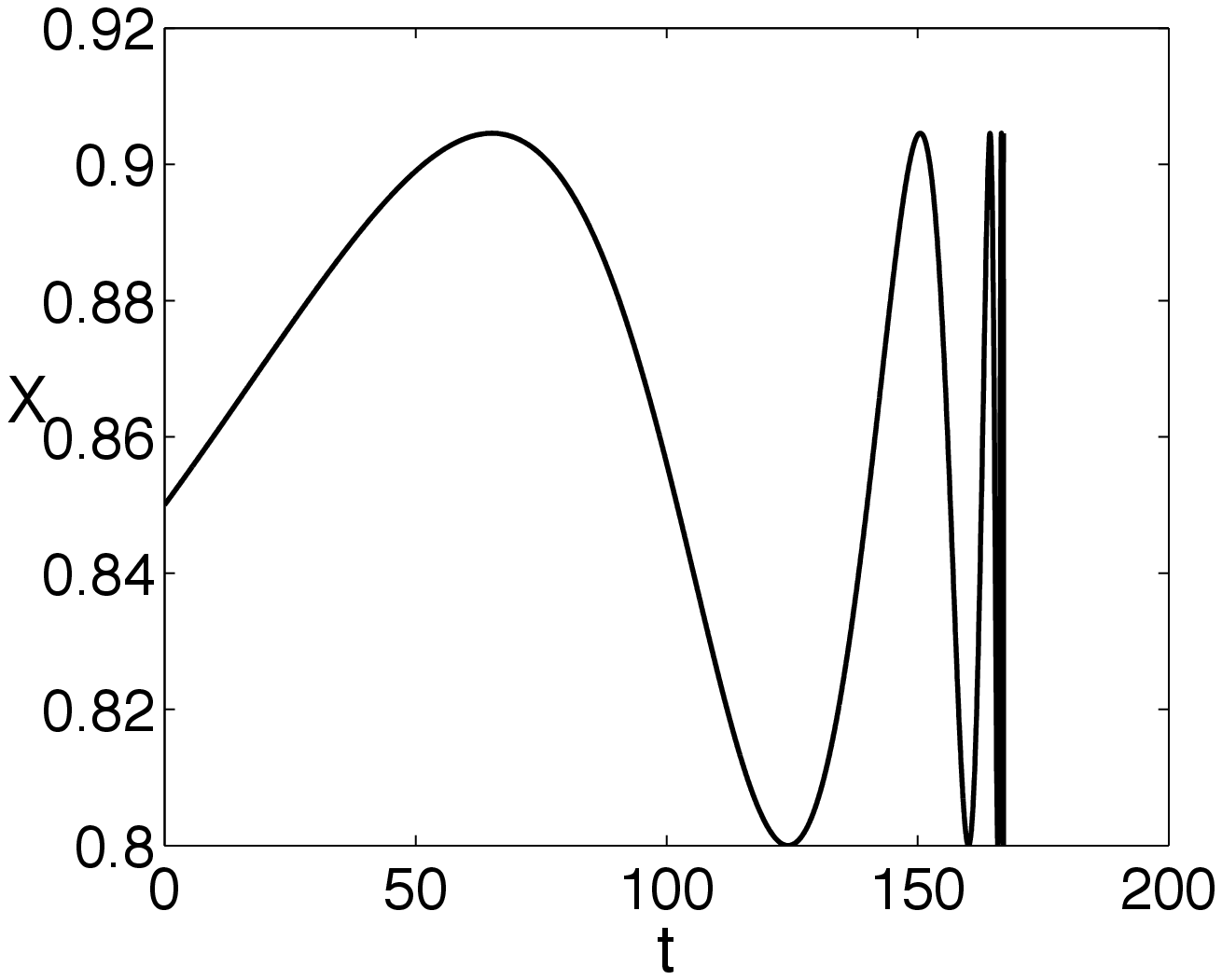} &
\includegraphics[width=5cm]{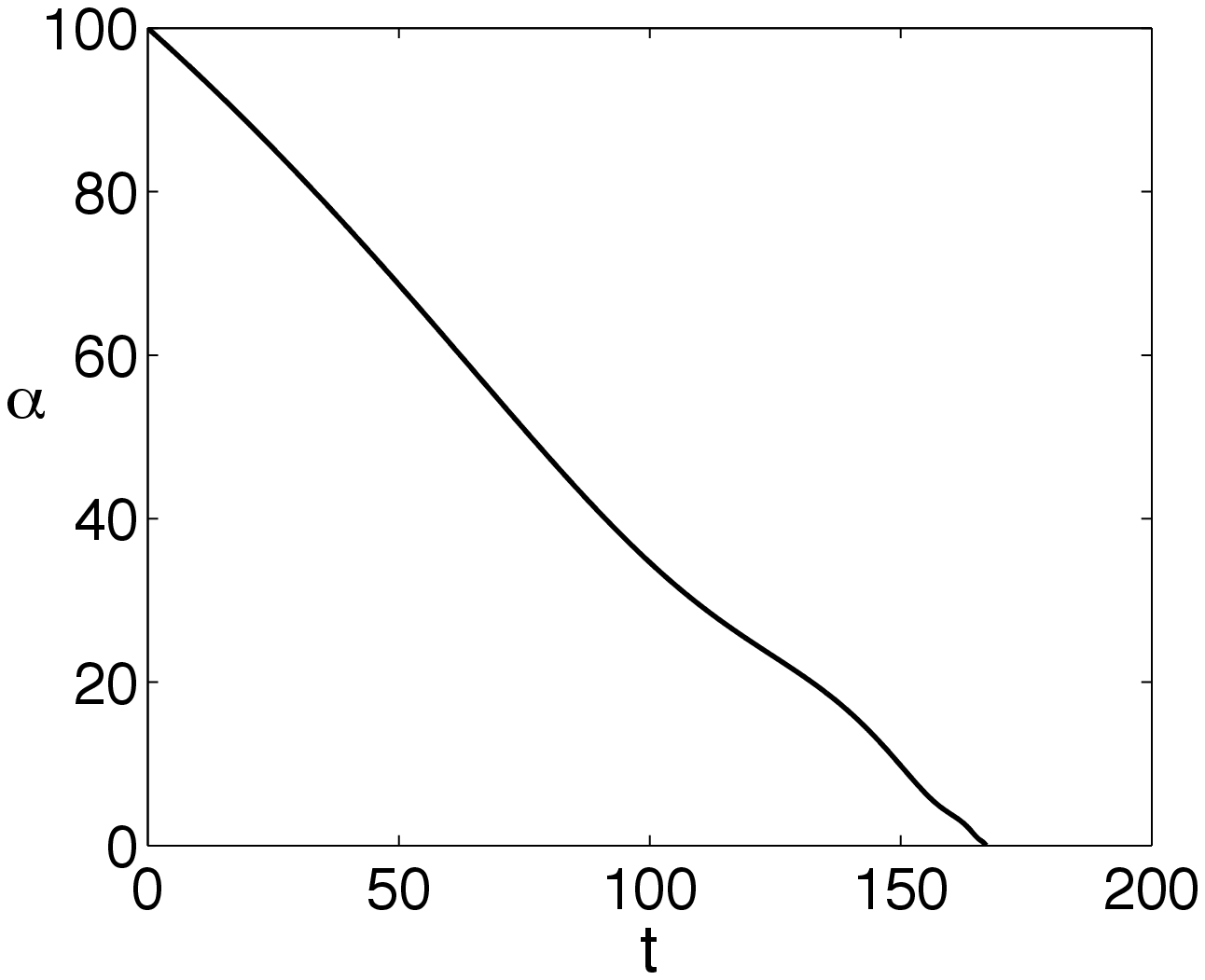} \\
\multicolumn{3}{c}{(b) $A=0.8$, $C=0.7$, $X_0=0.85$ and $\alpha_0=100$}\\
\\
\includegraphics[width=5cm]{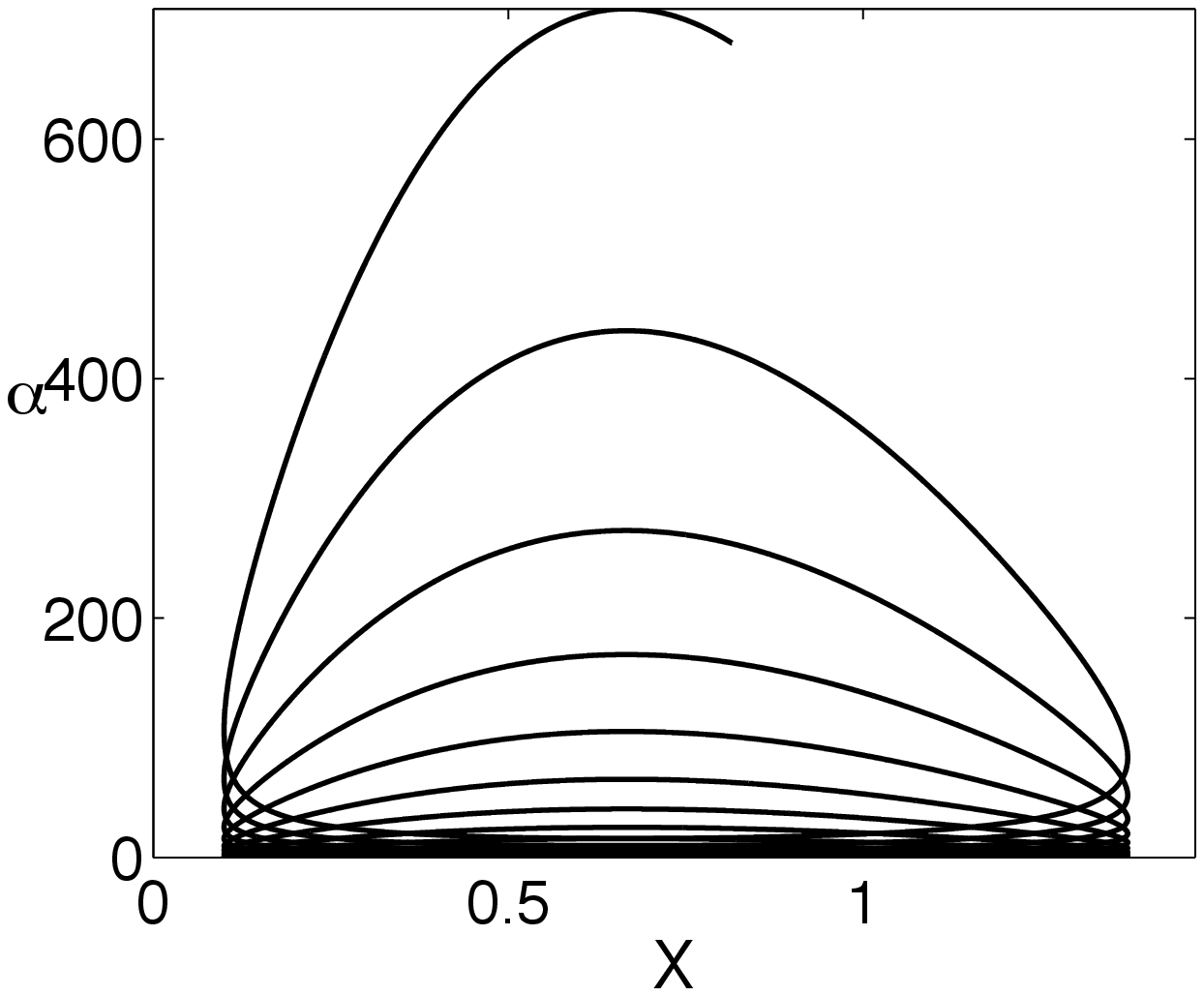} &
\includegraphics[width=5cm]{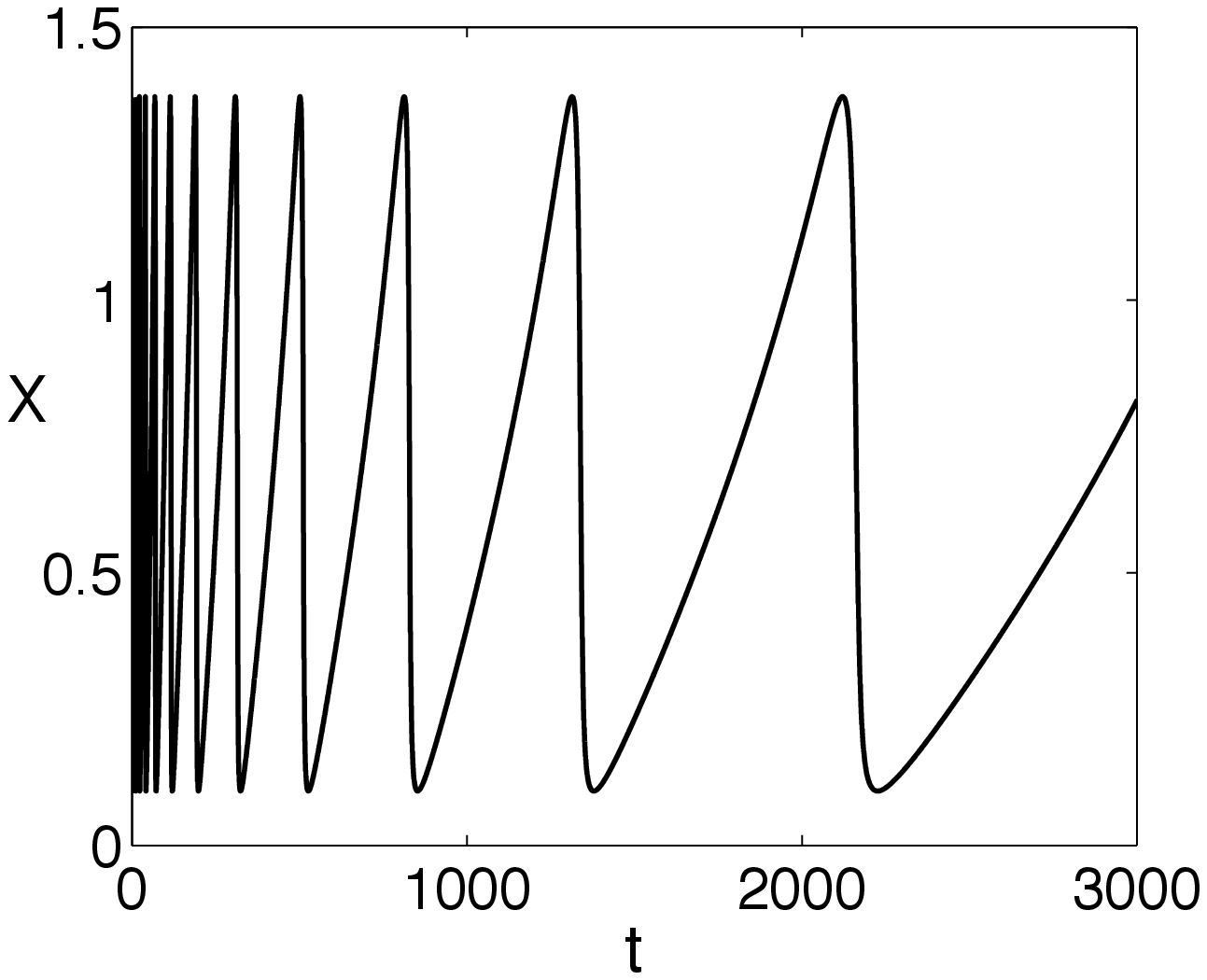} &
\includegraphics[width=5cm]{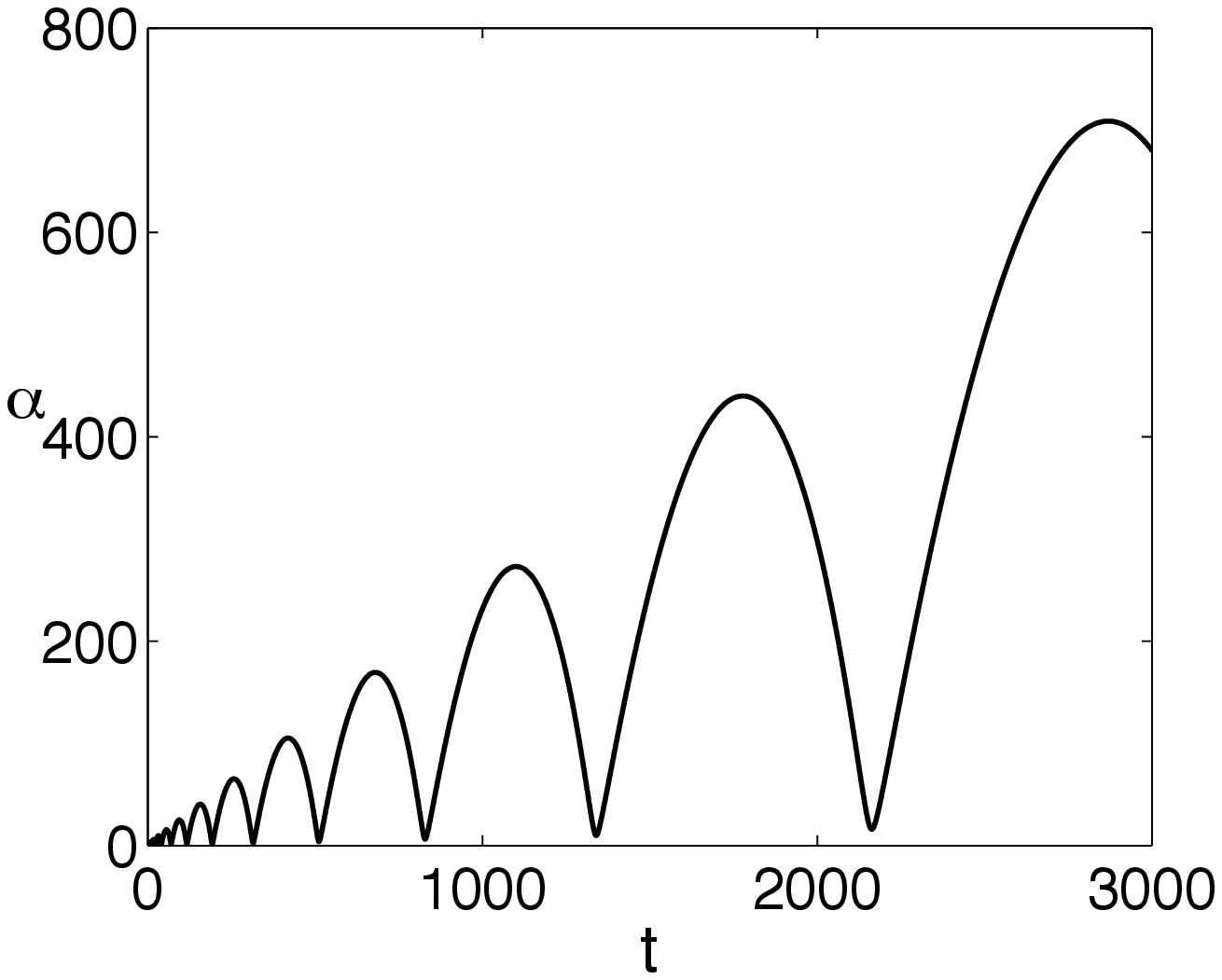} \\
\multicolumn{3}{c}{(c) $A=0.1$, $C=0.9$, $X_0=1$ and $\alpha_0=1$}\\
\\
\includegraphics[width=5cm]{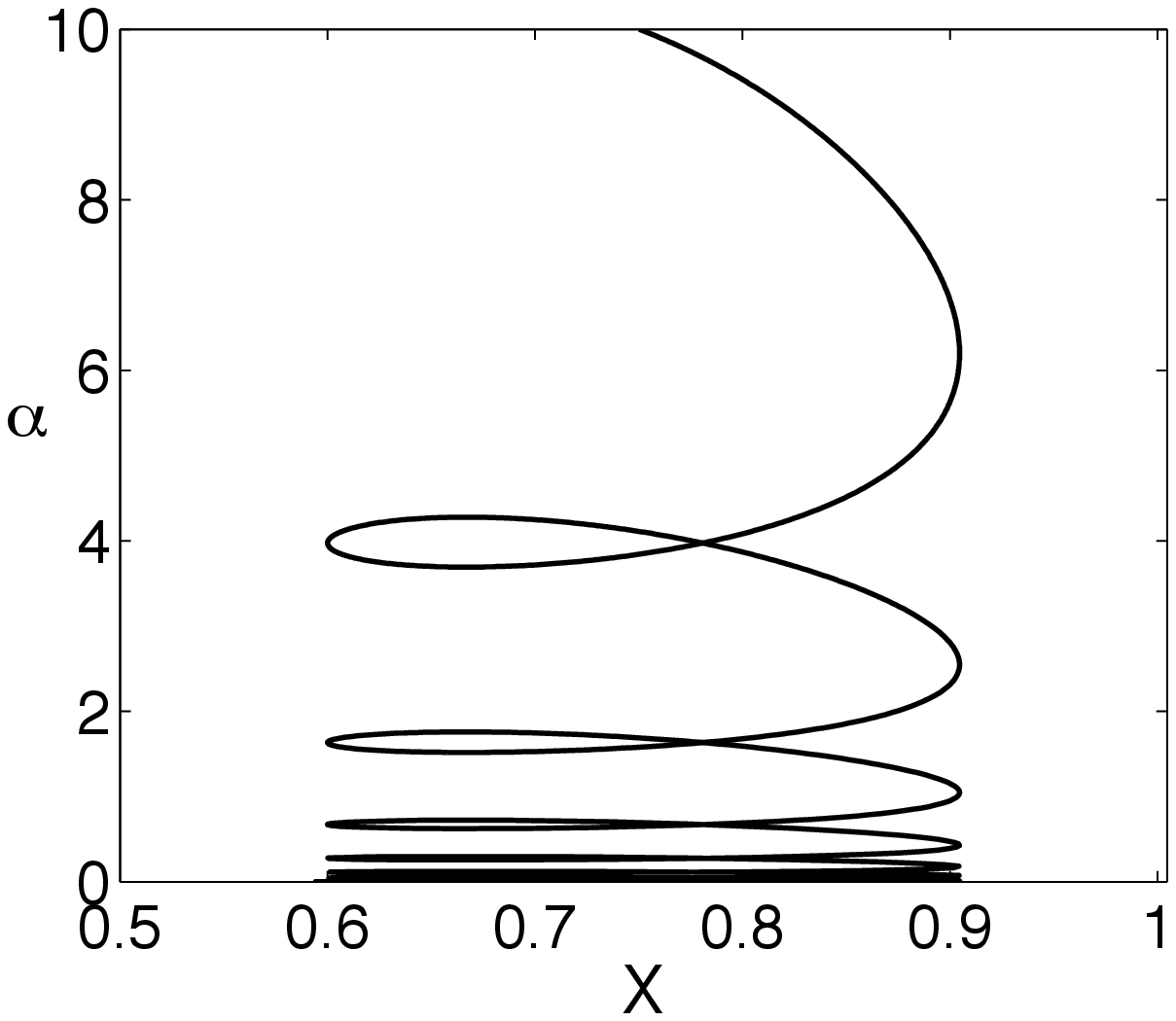} &
\includegraphics[width=5cm]{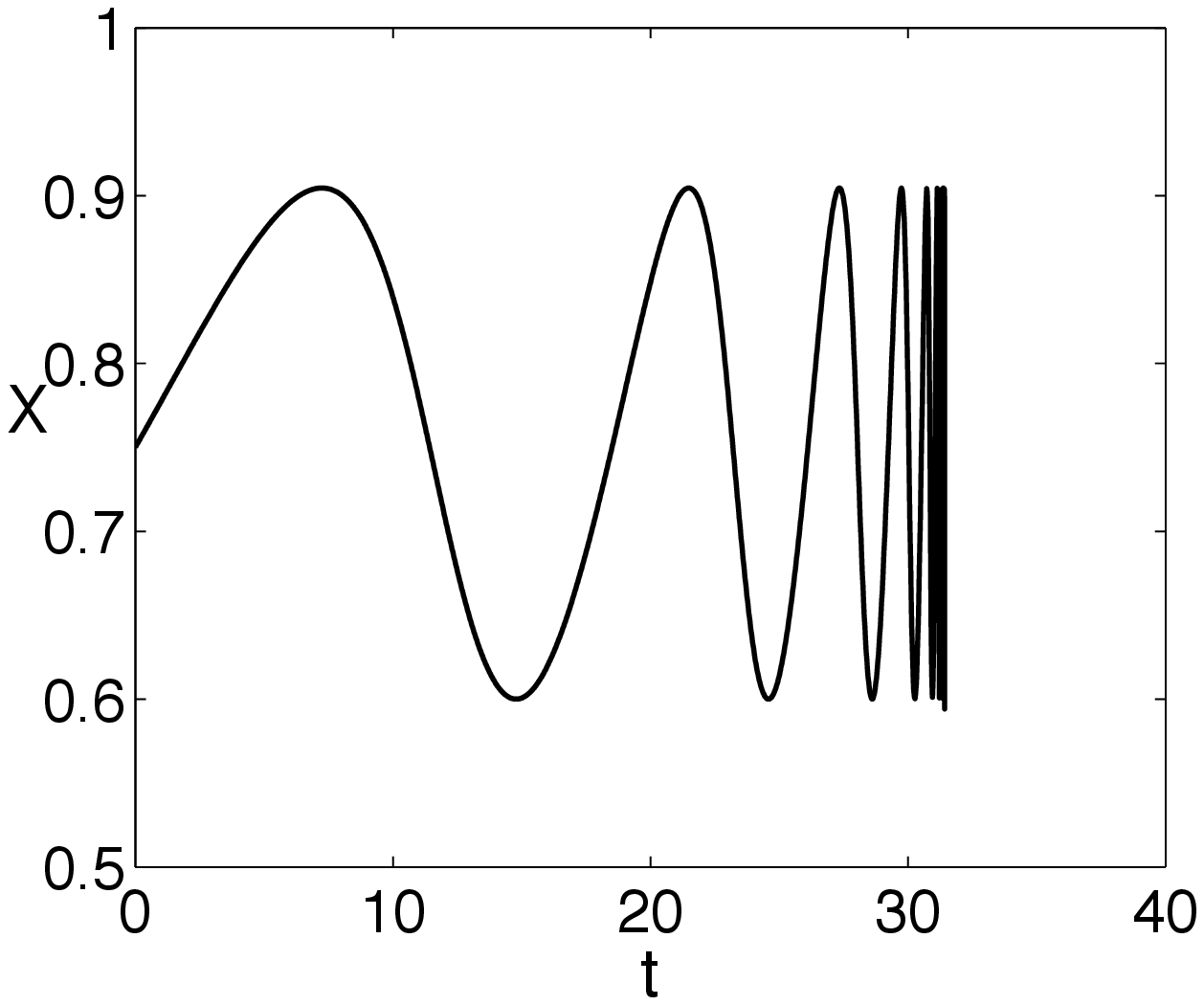} &
\includegraphics[width=5cm]{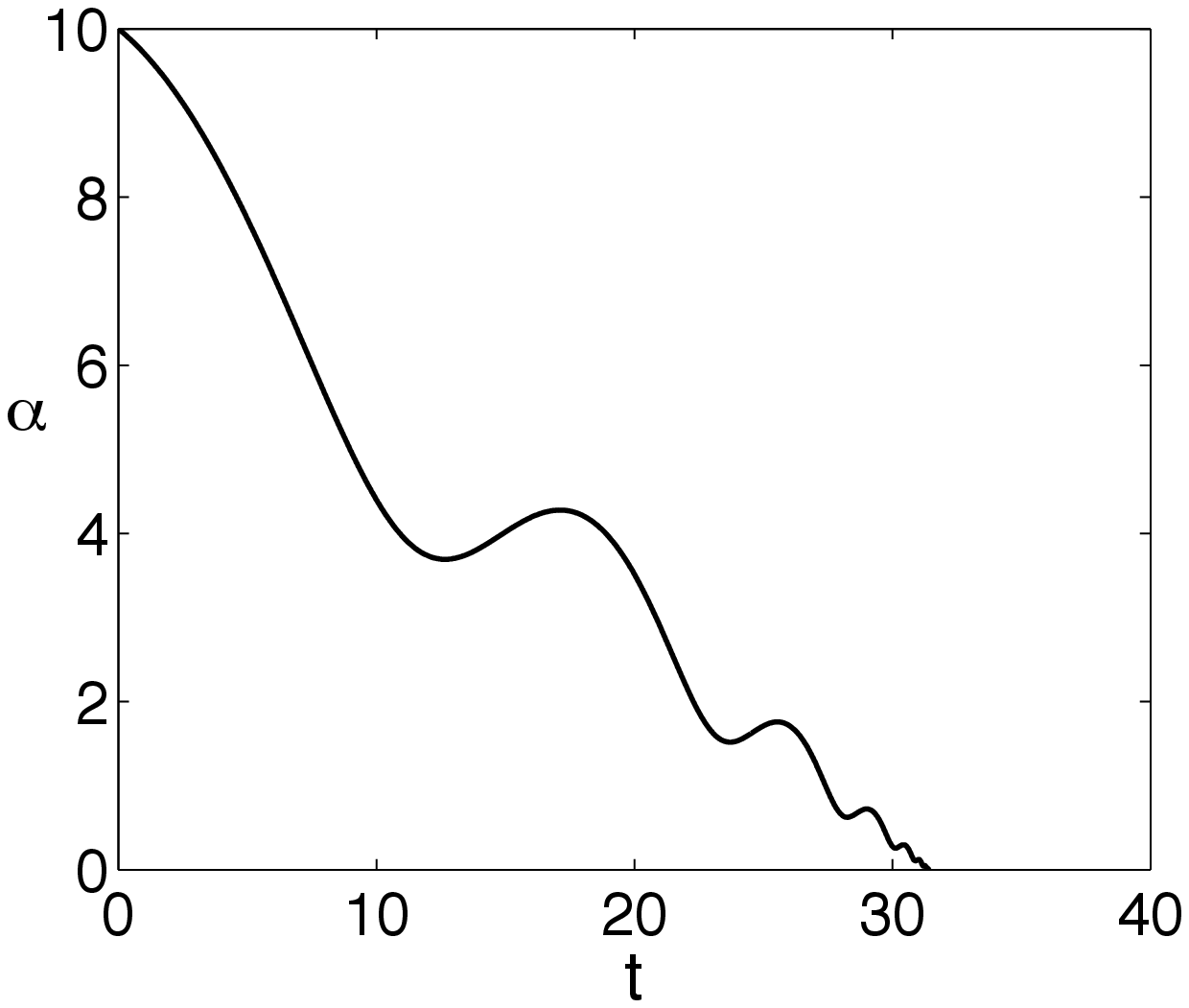} \\
\multicolumn{3}{c}{(d) $A=0.6$, $C=0.7$, $X_0=0.75$ and $\alpha_0=10$}
\end{tabular}
\caption{Illustration of the four possible trajectories in the phase plane $(X,\alpha)$ (left), evolution of $X$ (center) and evolution of $\alpha$ for four different configurations: (a) monotonic divergence, (b) monotonic convergence, (c) non-monotonic divergence and (d) non-monotonic convergence of the swimmers.}\label{fig:dynsyssol}
\end{center}
\end{figure}

As a followup to our theoretical analysis, we illustrate here the four different possible regimes obtained in \S\ref{theory}. These results are displayed in Fig.~\ref{fig:dynsyssol}.
If $A>2/3$ or $C<5/9$, the variation of $\alpha$ with time is monotonic, and can either be divergent (Fig.~\ref{fig:dynsyssol}a) or convergent (Fig.~\ref{fig:dynsyssol}b). If $A<2/3$ and $C>5/9$, then the variation of $\alpha$ over half an oscillation is not monotonic: $\dot{\alpha}$ changes of sign when $X=2/3$ corresponding to a minimum or maximum distance between the swimmers. This leads to a spiral shape of the trajectory in the plane $(X,\alpha)$, and non-monotonic divergence (Fig.~\ref{fig:dynsyssol}c) or convergence (Fig.~\ref{fig:dynsyssol}d) of the relative position between the swimmers.

\subsubsection{Finite time of collision}
We show here that in cases $(1)$ and $(4)$ discussed at the end of \S\ref{4cases}, the collision between the two swimmers occurs in a finite time. We consider case (4) for example. The time interval between two zeros of $\dot{X}$ is given by
\begin{equation}
\tau_n=t_{n+1}-t_n=\int_{X_{min}}^{X_{max}}\frac{\alpha(X)\dd X}{\sqrt{(X^2-A^2)(4C-4X+X^2)}}\leq\alpha_n \mathcal{T}(A,C),
\end{equation}
with 
\begin{equation}
\mathcal{T}(A,C)=\int_{X_{min}}^{X_{max}}\frac{\dd X}{\sqrt{(X^2-A^2)(4C-4X+X^2)}}\cdot
\end{equation}
Then we have
\begin{equation}
t_n=\sum_{k=0}^n\tau_k\leq \mathcal{T}\frac{1-\ee^{(n+1)\mathcal{G}}}{1-\ee^{\mathcal{G}}}\rightarrow\frac{\mathcal{T}}{1-\ee^{\mathcal{G}}},
\end{equation}
as $n\rightarrow\infty$ and the collision $\alpha_n=0$ happens at a finite time since $\mathcal{G}$ is negative. A bound on the finite collision time can be obtained in the same way for case $(1)$.

\subsubsection{Analysis of the system equilibrium}
Finally, we know from \S\ref{equilibrium} that the system has only one type of possible equilibrium. Using the notation from the current section, it corresponds to $A=X=2/3$ and $C=5/9$, $\alpha$ being arbitrary. This point is on the boundary between the collision and divergence domains on Fig.~\ref{fig:regimes}, as well as on the boundary $C=A-A^2/4$. One sees easily from Fig.~\ref{fig:regimes}, that for all values of $\alpha$ and for any value of the perturbation that does not leave $A$ and $C$ both unchanged, the system will move away from its equilibrium. This equilibrium is therefore nonlinearly unstable.

\subsubsection{Absolute displacement of the swimmers through hydrodynamic interactions}
In the previous sections, we have focused mostly on the relative motion of the two swimmers. The absolute motion is characterized by the evolution of $\rb_0$, defined as the middle point between the two swimmers. In the far-field approximation, we have computed the average velocity of this middle point as $\mathbf{v}_0$ in \eqref{average_abs}. Using the notations defined in sections \ref{redeq} and \ref{equilibrium}, $\mathbf{v}_0$ becomes
\[
\mathbf{v_0}=-\frac{3\mu\,A}{4r^2}\eb_z,
\]
where $\eb_z$ is a constant unit vector giving the direction of the relative distance between the swimmers. $A=\cos^2\theta_2-\cos^2\theta_1$ was shown to be a conserved quantity. The absolute motion of the swimmers therefore occurs along the same direction as the relative motion, and $\mathbf{v_0}$ does not change sign. In section \ref{conservation}, we relabeled the swimmers so that $A$ is a positive quantity. With this relabeling, the net motion of the swimmers occurs along $\eb_z$ in the direction of swimmer $1$ for pushers ($\mu>0$) and in the direction of swimmer $2$ for pullers ($\mu<0$). The net displacement velocity $|\mathbf{v}_0|$ scales like $1/r^2$, as expected from dipolar hydrodynamic interactions.

%%%% DISCUSSSION

\section{Discussion}
\label{discussion}
\subsection{Summary of results and biological relevance}

The work in this paper focuses on the hydrodynamic interaction of $N=2$ swimmers with circular trajectories, such as the spermatozoa of some marine  invertebrates \citep{goldstein77,riedel05}. This particular situation represents the simplest configuration in which the effects of hydrodynamic interactions can be studied without considering the full $N$-body problem with a large number of organisms. Indeed,  the confinement of the individual  trajectories allows the two swimming organisms to interact on a much longer time-scale than if they were swimming along straight lines. The two cells are assumed here to be spherical and identical, but the description of their swimming stroke is general, and the far-field interaction analysis is valid for an arbitrary stresslet tensor ({\it i.e.} an arbitrary force distribution at the swimmer surface). In general, the relative dynamics of the two cells is described by a dynamical system with nine degrees of freedom. 

In the far-field assumption, a separation of time scales occurs between the period of the intrinsic circular motion of the  swimmers  and the time over which hydrodynamic interactions have an order-one effect on their trajectories. As a result, the dynamical system is investigated using a multiple-scale analysis.
In particular, the  average motion resulting from the instantaneous interaction of the two swimmers is found to be strictly equivalent to the interaction of two modified stresslets, obtained as the stresslet for each swimmer averaged  over a period of its intrinsic motion (in other words, the time-averaged interaction between the swimmers is equal to the interactions between the time-averaged swimmers). Furthermore, the direction of the relative distance between the two swimmers is found to be independent of time, and the average problem was reduced to a four-dimensional dynamical system for the distance between the swimmers and the relative orientations of their rotation vectors.

We then proceed to a detailed mathematical analysis of the dynamical system. We show the existence of one type of equilibrium, which is linearly neutrally stable but nonlinearly unstable, and two types of rotational equilibria, which are both  linearly unstable with algebraic growth.  We then show  the existence of two conserved quantities, thereby allowing a reduction  to a two-dimensional dynamical system. We proceed to identify geometrical bounds on the dynamics, and we show that only two general long-time behaviors are possible:  Either the swimming cells swim away from each other, or they get closer from each other (until the far-field assumption breaks down). In these divergence and collision scenarios, the relative distance can either vary monotonically or can display oscillations, and the boundary between the two regimes is an unstable limit cycle.

The implication of our results for the dynamics of biological organisms is twofold. First, we show that there are no stable equilibria (in position or orientation) between the cells, a result which is true arbitrarily of the sign of the far-field flow field each cell is  generating (pushers or pullers). As a result, populations of cells are expected to always dynamically evolve, as is observed 
in experiments \citep{mendelson99,wu00,dombrowski04,kim04_diffusion,sokolov07,cisneros07} and modeling \citep{simha02,hernandez-ortiz05,aranon07,saintillan07,ishikawa_pedley_rheology07,ishikawa_pedley_diffusion07,
saintillan08,wolgemuth08,underhill08,ishikawa08,mehandia08} of cell populations,  with an intermittence  at the origin of the expression ``bacterial turbulence''.  The model system  studied in this paper allows us in particular to quantify rigorously  the rate at which the cells are being effectively repelled from, or attracted to each other, and to obtain all types of possible  swimming kinematics at $t\to\infty$. In addition, what this paper shows, is that hydrodynamic interactions leads to ``new" modes of swimming, meaning that the motion of each swimmer contains a component due to the presence of another cell which, over long times,  integrates to an order one change in its swimming kinematics. This is reminiscent of recent work showing that hydrodynamic interactions can impart motility to otherwise non-swimming  active bodies  \citep{alexander08,laugabartolo08}, and is relevant to the experimental observation that dense cell populations display different length, time and velocity scales than that of individual micro-organisms \citep{mendelson99,dombrowski04,sokolov07,cisneros07}.

\subsection{Modeling assumptions and possible extensions}\label{limitations}

The results in this paper were obtained under a number of simplifying assumptions, which we now discuss.

\subsubsection{Non-spherical swimmers}

We have first assumed  that the two swimmers are spherical, so that the rotation rate induced by the hydrodynamic interaction is equal to half the vorticity field created by the other swimmer. A corrective term of the same order  however appears as soon as the swimmer shape is not purely spherical. Analytic solutions have been obtained for ellipsoids \citep{jeffery1922,kimbook}. In this paper, we focus on the spherical case  as it provides the simplest system, and because it is a first good approximation of the shape of  spherical organisms using cilia or flagella whose effect on the induced rotation rate can be neglected if their size is small compared to the body of the swimmer. If the organism is not spherical,  a corrective term to the system, Eq.~(\ref{geneq}) must be added to account for the effect of anisotropy and local strain rate. In the case of an ellipsoidal swimmer, the induced rotation rate on swimmer $2$ is given by \citep{pedley92}
\begin{equation}\label{correctell}
\boldsymbol{\Omega}_{1\rightarrow 2}=\frac{1}{2}\boldsymbol\omega^{(1)}(\rb)+\beta_0\,\mathbf{p}_2\times\left(\mathbf{E}^{(1)}(\rb)\cdot\mathbf{p}_2\right),
\end{equation}
with $\rb=\rb_2-\rb_1$, $\boldsymbol\omega^{(1)}$ and $\mathbf{E}^{(1)}$ the vorticity field and strain rate tensor created by the motion of swimmer $1$, $\mathbf{p}_2$ the unit vector associated to the direction of the major axis of the ellipsoidal swimmer $2$ and $\beta_0=(c^2-1)/(c^2+1)$ with $c$ the ratio of major axis to minor axis of the ellipsoid, and measures the departure from the spherical case ($\mathbf{p}_2$ moves rigidly with the swimmer). A reasonable approximation is to consider that $\mathbf{p}_2=\eb_2$, {\it i.e.} the intrinsic swimming motion occurs in the direction of the major axis of the ellipsoid. The strain rate tensor is obtained from Eq.~(\ref{velindx}), and after substitution in Eq.~\eqref{correctell}, the induced rotation rate becomes 
\begin{align}
\boldsymbol{\Omega}_{1\rightarrow 2}=&\,\frac{\gamma\,\rb\times(\Sb^{(1)}\cdot\rb)}{r^5}\\
&+\gamma\alpha_0\left[\frac{5\left(\rb\cdot\Sb^{(1)}\cdot\rb\right)(\eb_2\cdot\rb)(\eb_2\times\rb)}{r^7}-\frac{(\eb_2\cdot\Sb^{(1)}\cdot\rb)(\eb_2\times\rb)}{r^5}-\frac{(\eb_2\cdot\rb)(\eb_2\times(\Sb^{(1)}\cdot\rb))}{r^5}\right].\nonumber
\end{align}

As pointed out above, the rotation rate now depends not only on the orientation of swimmer $1$ (through $\Sb^{(1)}$) but also on the orientation of swimmer $2$. In the limit of far-field interactions, the multiple-scale analysis of \S\ref{multscale2} is still valid and we can study the average motion of the swimmers as represented by their mean distance $\rb$ and the orientation of their intrinsic rotation vectors $\eb_i'$. However, the relative phase between the  circular motion of the swimmers (value between $0$ and $2\pi$) does not disappear in the averaged equations and acts as an additional arbitrary parameter.

\subsubsection{Validity of the far-field approximation and regularization}

When the two swimmers are not far enough from each other, two of our assumptions successively break down.
Firstly, the separation of time scales is no longer valid when the time-scale associated with the intrinsic rotation of each swimmer is no longer much smaller than the time scale associated with the hydrodynamic interaction. In that case, the multiple scale analysis of \S\ref{multscale}  breaks down, and one needs instead to consider the full coupled equations, Eq.~\eqref{geneq}.

Secondly, when the swimmers get close to each other, the description of hydrodynamic interactions as being dominated by their far-field limit is no longer valid, and the following three  terms need to be considered: 
(a) Higher-order corrections in the velocity and vorticity field created by a swimmer in Eq.~\eqref{farfieldvel} (flows with $r^{-3}$ decay such as force-quadrupoles, source-dipoles; flows with  $r^{-4}$ decay etc.). 
(b) Higher-order corrections in the induced velocity on a swimmer whose size is no longer negligible compared to the characteristic length-scale of the local flow. For a sphere, the exact correction is given by F\`axen's law. Generalized exact formulae can be obtained for ellipsoids \citep{jeffery1922,lamb1932}. For arbitrary shape, general frameworks have been studied allowing the computation of the successive corrective terms \citep{brenner1964,liron1992};
(c) Higher-order corrections due to the two-way coupling: Swimmer $1$ creates a flow field that influences swimmer $2$. But the presence of swimmer $2$, modifies this flow field (even if swimmer $2$ was not swimming) which also induces a correction on the velocity of swimmer $1$. 

These three contributions are negligible for large distances but can become dominant at intermediate distance or in near-field interactions. In particular, we observed previously that the far-field behavior can lead to collisions between the swimmers (when $\alpha$ or $r$ go to zero) as the hydrodynamic interaction terms in Eq.~(\ref{geneq}) are attractive for particular relative orientations of the swimmers, regardless of their relative distance. Moreover, these attractive interactions also have a diverging amplitude as $r\rightarrow 0$. Obviously, the far-field approximation is violated when the distance becomes small, and the higher order corrections discussed above must be included to account for regularizing  forces that arise at intermediate or short distances. In an effort to remain  general, one could attempt to introduce some empirical short distance corrections to reduce the attractive terms in the near-field (in the form of exponential or power-laws regularization at small $r$ for example), but these are not based on physical principles. For distances between the swimmers much smaller than the  typical size of each swimmer, lubrication theory can be used but, in the intermediate distance range numerical simulation is necessary \citep{ishikawa06}.

In general, if one is interested in intermediate or short range interactions, a knowledge of the detailed swimmer geometry and  propulsion method is necessary, as is the case for  spherical squirmers \citep{ishikawa06,ishikawa_pedley_rheology07,ishikawa_pedley_diffusion07} or dumbbell-like model organisms \citep{hernandez-ortiz05,gyria2009}. Squirmers maintain a spherical shape at all time and generate motion by tangential displacement of their surfaces. They are generally thought as a good approximation for spherical swimmer using ciliary propulsive schemes, the spherical shape of the swimmer corresponding to the envelope of the cilia in that case. The squirmer formulation has the advantage that an analytic solution exists for the velocity field created. Using F\`axen's law, an exact system for spherical swimmers can then be obtained. Analytic solution of the multi-body problem is however not possible in general and such a system must be solved numerically. Both squirmers and dumbbell-like organisms are simple approximations of real swimmers, and considerations of the detailed geometry of the swimmer often lead to a trade-off between accuracy in the biophysical description of real organisms and simplified representations to allow an easier mathematical or numerical treatment.

\section*{Acknowledgments} 
This work was funded in part by the US National Science Foundation (grants CTS-0624830 and CBET-0746285 to Eric Lauga).

\end{document}